\documentclass[fleqn,usenatbib]{mnras}

\usepackage{newtxtext,newtxmath}

\usepackage[T1]{fontenc}
\usepackage{ae,aecompl}

\usepackage{graphicx}
\usepackage{amsmath}
\usepackage{amssymb}
\usepackage{float}

\title[Low-$z$ BAO from the Reconstructed 6dFGS]{Low Redshift Baryon Acoustic Oscillation Measurement from the Reconstructed 6-degree Field Galaxy Survey}

\author[P. Carter et al.]{Paul Carter$^{1}$\thanks{E-mail: paul.carter1@port.ac.uk},
Florian Beutler$^{1, 2}$,
Will J. Percival$^{3, 4, 1}$,
Chris Blake$^{5, 8}$,
Jun Koda$^{6, 7, 8}$,
\newauthor Ashley J. Ross$^{9}$
\\
$^{1}$Institute of Cosmology \& Gravitation, University of Portsmouth, Dennis Sciama Building, Portsmouth, PO1 3FX, UK\\
$^{2}$Lawrence Berkeley National Lab, 1 Cyclotron Rd, Berkeley CA 94720, USA\\
$^{3}$Department of Physics and Astronomy, University of Waterloo, 200 University Ave W, Waterloo, ON N2L 3G1, Canada\\
$^{4}$Perimeter Institute for Theoretical Physics, 31 Caroline St. North, Waterloo, ON N2L 2Y5, Canada\\
$^{5}$Centre for Astrophysics \& Supercomputing, Swinburne University of Technology, PO Box 218, Hawthorn, VIC 3122, Australia\\
$^{6}$Dipartimento di Matematica e Fisica, Universit\'a degli Studi Roma Tre, via della Vasca, Navale 84, I-00146 Roma, Italy\\
$^{7}$INFN Sezione di Roma 3, Via della Vasca Navale 84, Rome 00146, Italy\\
$^{8}$ARC Centre of Excellence for All-sky Astrophysics (CAASTRO)\\
$^{9}$Center for Cosmology and AstroParticle Physics, The Ohio State University, Columbus, OH 43210, USA
}

\date{Accepted XXX. Received YYY; in original form ZZZ}

\pubyear{2018}

\begin{document}
\label{firstpage}
\pagerange{\pageref{firstpage}--\pageref{lastpage}}
\maketitle

\begin{abstract}
Low redshift measurements of Baryon Acoustic Oscillations (BAO) test the late time evolution of the Universe and are a vital probe of Dark Energy. Over the past decade both the 6-degree Field Galaxy Survey (6dFGS) and Sloan Digital Sky Survey (SDSS) have provided important distance constraints at $z < 0.3$. In this paper we re-evaluate the cosmological information from the BAO detection in 6dFGS making use of HOD populated COLA mocks for a robust covariance matrix and taking advantage of the now commonly implemented technique of density field reconstruction. For the 6dFGS data, we find consistency with the previous analysis, and obtain an isotropic volume averaged distance measurement of $D_{V}(z_{\mathrm{eff}}=0.097) = 372\pm17(r_{s}/r_{s}^{\mathrm{fid}})\,\mathrm{Mpc}$, which has a non-Gaussian likelihood outside the $1\sigma$ region. We combine our measurement from both the post-reconstruction clustering of 6dFGS and SDSS MGS offering the most robust constraint to date in this redshift regime, $D_{V}(z_{\mathrm{eff}}=0.122)=539\pm17(r_{s}/r^{\mathrm{fid}}_{s})\,\mathrm{Mpc}$. These measurements are consistent with standard $\Lambda\mathrm{CDM}$ and after fixing the standard ruler using a Planck prior on $\Omega_{m}h^{2}$, the joint analysis gives $H_{0}=64.0\pm3.5\,\mathrm{kms}^{-1}\mathrm{Mpc}^{-1}$. In the near future both the Taipan Galaxy Survey and the Dark Energy Spectroscopic Instrument (DESI) will improve this measurement to $1\%$ at low redshift.
\end{abstract}


\section{Introduction}
\label{sec:intro}

Utilisation of the baryon acoustic peak feature measured in the two-point statistics of redshift surveys, has been integral in the advancement to precision cosmology. Within the framework of the concordance $\Lambda$CDM model the initial matter perturbations were seeded through quantum fluctuations during an early inflationary epoch. Following this the Universe existed in a plasma state, in which radiation and baryonic matter are strongly coupled through the process of Thompson scattering.

Interplay between gravitational attraction and radiation pressure introduces acoustic oscillations in the primordial photon-baryon fluid. At $z \approx 1100$ the background temperature of the Universe becomes comparable to the ionisation energy of electrons and recombination occurs. The mean free path of photons becomes greater than the Hubble distance and they therefore decouple. This process leaves the baryonic matter distributed in overdensities in a surrounding spherical shell. These shells have a co-moving radius which corresponds to the sound horizon, $r_{s} \sim 150$Mpc, at the surface of last scattering. Through mutual gravitational interaction the dark matter component grows with the baryonic component to emulate this feature \citep{2007ApJ...664..660E}. Galaxies form in regions of overdensity and are biased tracers of the underlying dark matter field on large scales.

Making use of the two-point correlation function $\xi$ it is possible to see the effect of BAO as a peak, centered about separation $r_{s}$. Alternatively the power spectrum gives equivalent information in Fourier space, with the signal translating to oscillations in amplitude with wavenumber. Measurements of the apparent scale of the BAO features, offer a robust standard ruler of the distance to the measured galaxy population's effective redshift. Measurements of this feature at different redshifts allow for a test of the general cosmological model, especially inference regarding the equation of state of Dark Energy and spatial curvature.

The BAO peak is a well studied probe of cosmology, first detected in both the initial Sloan Digital Sky Survey (SDSS; \citealt{2000AJ....120.1579Y}) and 2-degree Field (2dFGRS; \citealt{2001MNRAS.328.1039C}) galaxy redshift surveys \citep{2001MNRAS.327.1297P, 2005ApJ...633..560E, 2005MNRAS.362..505C}. Subsequently the BAO peak has been detected in later SDSS data releases \citep{2010MNRAS.401.2148P, 2010ApJ...710.1444K}, 6dFGS \citep{2011MNRAS.416.3017B}, WiggleZ \citep{2011MNRAS.415.2892B}, BOSS (LOWZ and CMASS) \citep{2017MNRAS.470.2617A}, eBOSS luminous red galaxies (LRGs) \citep{2017arXiv171208064B} and quasars (QSOs) \citep{2018MNRAS.473.4773A} and a higher redshift detection using Ly-$\alpha$ forest measurements in BOSS \citep{2013JCAP...04..026S, 2014JCAP...05..027F, 2015A&A...574A..59D}. These measurements have constructed a distance ladder that spans from $z = 0$ out to $z \sim 0.8$ using conventional galaxy redshift surveys, $z \sim 1.5$ through eBOSS QSO and to $z \sim 2.3$ when including Ly-$\alpha$.

In recent analyses of the BAO peak, a method of density field reconstruction has been employed. \citet{2007ApJ...664..675E} proposed that, as the bulk flows that smear the acoustic peak are sourced from the density field potential itself, the galaxy map can itself be used to estimate the displacement field. Removal of these shifts has been shown to reduce the damping of the BAO and increase the $S/N$ of this feature. This increased $S/N$ results from higher order statistics which have been moved back into linear fluctuations \citep{2015PhRvD..92l3522S}.

Density field reconstruction has been applied in recent work including SDSS \citep{2012MNRAS.427.2132P,2015MNRAS.449..835R}, WiggleZ \citep{2014MNRAS.441.3524K} and throughout BOSS \citep{2017MNRAS.470.2617A}. These studies use either a perturbation theory based approach that relies on the finite difference method \citep{2009PhRvD..79f3523P, 2009PhRvD..80l3501N}, or an alternative FFT-based iterative algorithm \citep{2014MNRAS.445.3152B, 2015MNRAS.453..456B}.

This paper explores the application of density field reconstruction to 6dFGS following from the initial detection of the BAO peak by \citet{2011MNRAS.416.3017B}. Aside from providing post-reconstruction constraints, the paper also improves on the analysis of errors, using high fidelity COLA-based mock catalogues \citep{2016MNRAS.459.2118K} to provide more robust covariance matrices than the lognormal approach adopted previously. We also combine with the \citet{2015MNRAS.449..835R} lowest redshift measurement, using the SDSS-II Main Galaxy Sample (MGS). This combined low-redshift measurement is useful in providing further constraining power on a direct measurement of $H_{0}$ for cosmological tests of dark energy and curvature. Also this offers an independent test of the Hubble constant at the redshift of the supernovae surveys, for which there is currently tension with other BAO and CMB studies.

This paper is organised as follows: Section~\ref{sec:dataset} offers an overview of the final 6dFGS dataset and describes the specifications of the catalogue used. Section~\ref{sec:mock} outlines the use of halo occupation distribution (HOD) modelling to provide a basis on which to develop high-fidelity COLA-based mock catalogues. Section~\ref{sec:over} gives a summary of the clustering statistic measurement made during this work and the formalism used for density field reconstruction. Section~\ref{sec:model} provides the model used and constraints extracted both before and after reconstruction. Following this a number of tests regarding the mock population and robustness of results are conducted in Section~\ref{sec:tests}. Section~\ref{sec:cosmo} provides a joint analysis with SDSS DR7 MGS data and the cosmological interpretation of theses results. Finally, our conclusions are given in Section~\ref{sec:concl}.

Our analysis uses a fiducial flat $\Lambda$CDM model cosmology with parameters $\Omega_{m}^{\mathrm{fid}}=0.31$, $\Omega_{\Lambda}^{\mathrm{fid}}=0.69$, $h^{\mathrm{fid}}=0.67$ where $H_{0} = 100\,h$km s$^{-1}$ Mpc$^{-1}$ and a fiducial sound horizon size at the drag epoch is $r_{s}(z_{d})=147.5\,\mathrm{Mpc}$.

\section{The 6dF Galaxy Survey}
\label{sec:dataset}

\begin{figure}
	\includegraphics[width=\columnwidth]{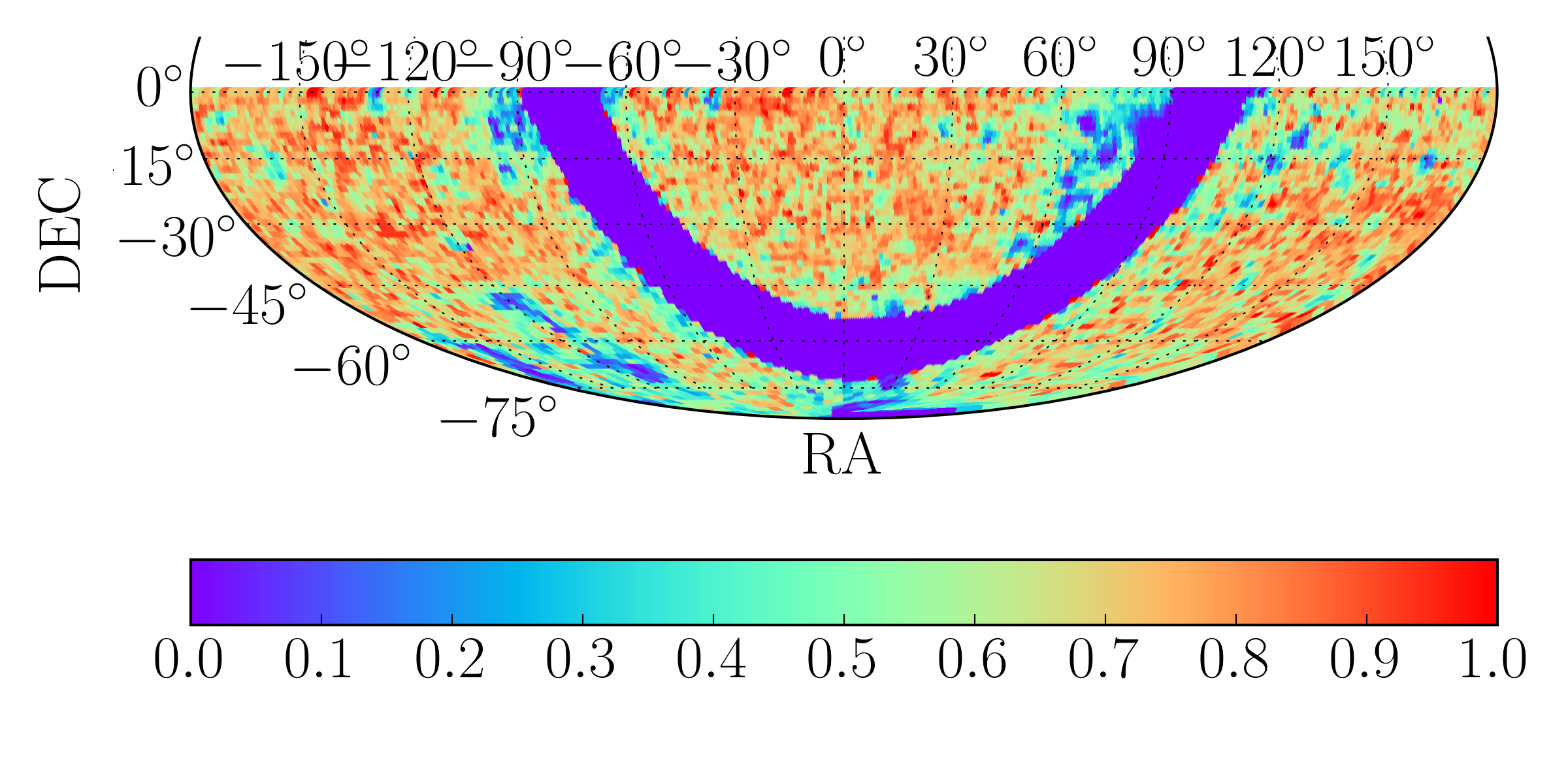}
    \caption{The sky coverage of the 6dFGS K-band sample, the colour of each cell corresponds to the completeness in that region. This total completeness constitutes a combination of sky and magnitude completeness of galaxies.}
    \label{fig:figure1a}
\end{figure}

The 6dF Galaxy Survey\footnote{The data, randoms and mock catalogues used during this work are available at: \href{http://www.6dfgs.net/downloads/6dFGS_Recon_Files.tar.gz}{http://www.6dfgs.net/downloads/6dFGS\_Recon\_Files.tar.gz}} (6dFGS; \citealt{2009MNRAS.399..683J}) combines peculiar velocity and redshifts for galaxies covering almost the entire southern hemisphere. The survey was conducted between 2001 and 2006 using the 6-degree Field multi-fibre instrument on the UK Schmidt Telescope and covers $17,000\,\mathrm{deg}^{2}$ of the sky with a median redshift $0.053$. The sample selected for this work is the same as that used in \citet{2011MNRAS.416.3017B}, selected with a magnitude cut of $K < 12.9$ and imposing a cut for regions with $<60\%$ completeness \citep{2006MNRAS.369...25J}. This gives a catalogue of 75,117 galaxies. In Figure~\ref{fig:figure1a} the sky coverage and total completeness distribution of galaxies in this final catalogue are presented. Further details regarding the structure and systematics in the full 6dFGS can be found in \cite{2009MNRAS.399..683J}.

\begin{figure}
	\includegraphics[width=\columnwidth]{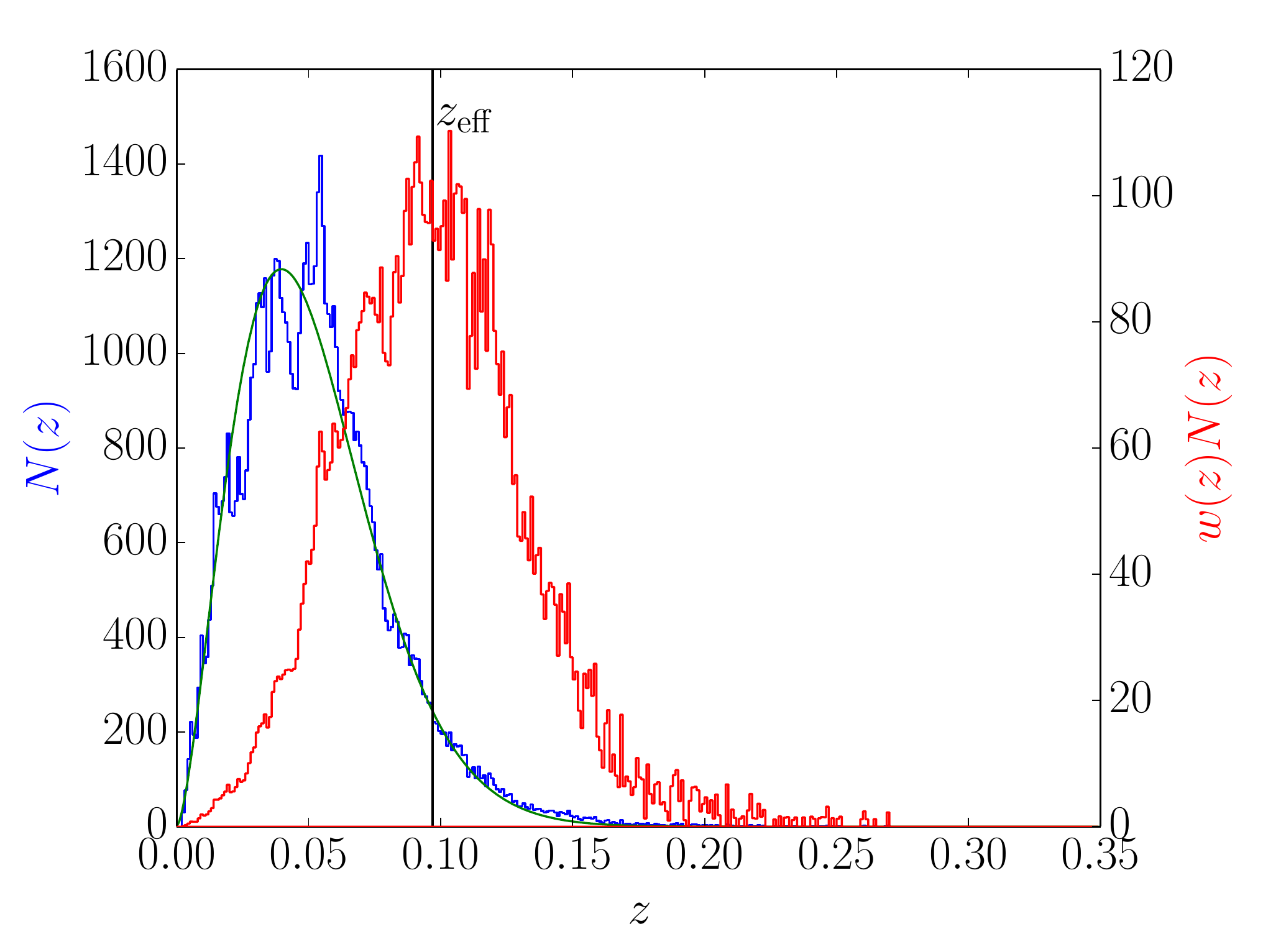}
    \caption{The unweighted (blue) and FKP weighted (red) redshift distribution of 6dFGS, with the best fitting model of $N_{\mathrm{fit}}(z)$ to the unweighted distribution. The effective redshift of the 6dFGS in given as the black vertical line.}
    \label{fig:figure2}
\end{figure}

Throughout this work galaxies have $S/N$ (FKP) weights applied \citep{1994ApJ...426...23F}. These weights $w_{\mathrm{FKP}}(z) = 1/(1+n(z)P_{0})$, optimise between sample variance and shot noise on the $k$-scale which is specified by $P_{0}$ the amplitude of the power spectrum and $n(z)$ is the number density. 6dFGS has a high density gradient with redshift, and therefore this weighting scheme has a strong impact on the analysis.
\noindent We chose to use $P_{0} \sim 10,000\,h^{-3}$Mpc$^{3}$ approximately the amplitude where the BAO signal peaks, $k_{\mathrm{eff}} \sim 0.15\,h$Mpc$^{-1}$ \citep{2007ApJ...665...14S}. A fit to the unweighted redshift distribution is used to calculate $n(z)$ for the weighting. The form of this model is $N_{\mathrm{fit}}(z)=Az^{\gamma}\exp[-(z/z_{p})^{\gamma}]$ with the best fit parameters $A=456500$, $z_{p}=0.03967$ and $\gamma=1.5369$ \citep{2009MNRAS.399..683J}. The redshift distribution both unweighted and once the FKP weighting scheme has been applied are shown in Figure~\ref{fig:figure2}. The effective redshift of the total weighted sample is $z_{\mathrm{eff}} = 0.097$ where the effective redshift is defined as,
\begin{equation}
    z_{\mathrm{eff}} = \frac{\sum\limits_{i = 0}^{N_{g}}z_{i}w_{i}(z)}{\sum\limits_{i = 0}^{N_{g}}w_{i}(z)}.
	\label{eq:equ1}
\end{equation}

\section{COLA-based Mock Catalogues}
\label{sec:mock}

Mock catalogues are produced based on numerical simulations and populated using a Halo Occupation Distribution (HOD) model. We construct our covariance matrix from 600 of these realisations which simulate the survey volume.

\subsection{COLA Simulations}
\label{sec:mock1}

In an ideal world the underlying dark matter haloes would be generated using $N$-body simulations, to encapsulate fully the non-linear regime, however this is computationally expensive. Instead we make use of the COmoving Lagrangian Acceleration (COLA) method (\citealt{2013JCAP...06..036T}, \citealt{2015A&C....12..109H}, \citealt{2016MNRAS.459.2118K}) to produce our 6dFGS mock catalogues.
\par 
The COLA method allows for an increase in efficiency by combining $N$-body simulations with 2nd-order Lagrangian Perturbation Theory (2LPT). By allowing 2LPT to solve the large scale distribution exactly and only using time-stepping for the small scales, COLA outperforms classical $N$-body simulations in terms of speed of computation by orders of magnitude.
\par To run the COLA simulations we used 432 cores and 8GB of memory per core on the Raijin supercomputer at the Australian National Computational
Infrastructure (NCI), with each run taking $\sim 45$ minutes. The mocks produced consist of $(1728)^{3}$ particles in boxes with $1.2 h^{-1}\,\mathrm{Gpc}$ on each side. The number of time steps of the COLA simulation is 20, down to $z=0$. The mass resolution of the mock catalogues is $2.8\times 10^{10}\,h^{-1}M_{\odot}$ and a friends-of-friends (FoF) finder is used to locate haloes that consist of a minimum of $32$ dark matter particles. The initial fiducial cosmology used in these mocks is $\Omega_{m}=0.3$, $\Omega_{b}=0.0478$, $h=0.68$, $\sigma_{8}=0.82$ and $n_{s}=0.96$. To ensure our simulation matches the data catalogue, snapshots at $z = 0.1$ were used to build the mock surveys, close to the effective redshift of the data.
\par The fundamental mass resolution of the mock catalogue does mean that we are not able to replicate the low mass haloes observed at low redshift. This will lead to a discrepancy between data and mock at $z < 0.05$, in the HOD population. This does not impact the BAO analysis on large scales; the reduction of number density at low redshift from a lack of these low mass galaxies only contributes to a reduction in effective volume of $1.0\%$.

\subsection{Halo Occupation Distribution Model}
\label{sec:mock2}

\par As is common in HOD modelling, we separate the clustering contribution between massive central and satellite galaxies. The probability of a dark matter halo of mass $M$, hosting each type is respectively \citep{2005ApJ...633..791Z},
\begin{equation}
\left<N_{C}(M)\right> = \frac{1}{2}\left[1 + \textrm{erf}\left(\frac{\log_{10}(M)-\log_{10}(M_{\textrm{min}})}{\sigma_{\log M}}\right)\right],
\label{eq:equ2}
\end{equation}
and,
\begin{equation}
\left<N_{S}(M)\right> = \left(\frac{M-M_{\mathrm{min}}}{M_{1}}\right)^{\alpha},
\label{eq:equ3}
\end{equation}
in which $M_{\mathrm{min}}$ is the minimum dark matter halo mass that can host a central galaxy, $M_{1}$ is the mass of a halo that on average contains one additional satellite member and $\sigma_{\log M}$ (\citealt{2011ApJ...736...59Z}) allows for a gradual transition from $\left<N_{C}(M)\right> = 0$ to $\left<N_{C}(M)\right> = 1$. When assigning the satellites to the halo, peculiar velocities are randomly selected based on the virial velocity of the halo assuming a spherical \cite{1996ApJ...462..563N} profile. As a dark matter halo can only host a satellite if it already has a central galaxy, the total halo occupation is,
\begin{equation}
\left<N_{t}(M)\right> = \left<N_{C}(M)\right>\left(1 + \left<N_{S}(M)\right>\right).
\label{eq:equ4}
\end{equation}
\par The halo model accounts separately for clustering between galaxies in the same halo (1-halo term $\xi_{1h}(r)$) and those in different haloes (2-halo term $\xi_{2h}(r)$), which contribute to the correlation function on small scales and large scales respectively. The total correlation function is just the superposition of these terms $\xi(r) = \xi_{1h}(r) + \xi_{2h}(r)$.

\subsubsection{Fitting of the HOD model}
To replicate the data, both the redshift distribution $n(z)$ and projected correlation function $w_{\mathrm{p}}(r_{\mathrm{p}})$,
\begin{equation}
w_{\mathrm{p}}(r_{\mathrm{p}}) = 2\int\limits^{\pi_{\mathrm{max}}}_{0}d\pi \xi(r_{\mathrm{p}}, \pi),
\label{eq:equ5}
\end{equation}
are used to measure the HOD parameters. $r_{\mathrm{p}}$ and $\pi$ correspond to the separation between galaxies perpendicular and parallel to the line-of-sight respectively. When integrating along the line-of-sight we use $\pi_{\mathrm{max}}=50h^{-1}\mathrm{Mpc}$, producing a statistic that has negligible contribution from the Fingers-of-God (FoG) redshift space distortion (RSD) effect.

\begin{figure*}
	\includegraphics[width=2\columnwidth]{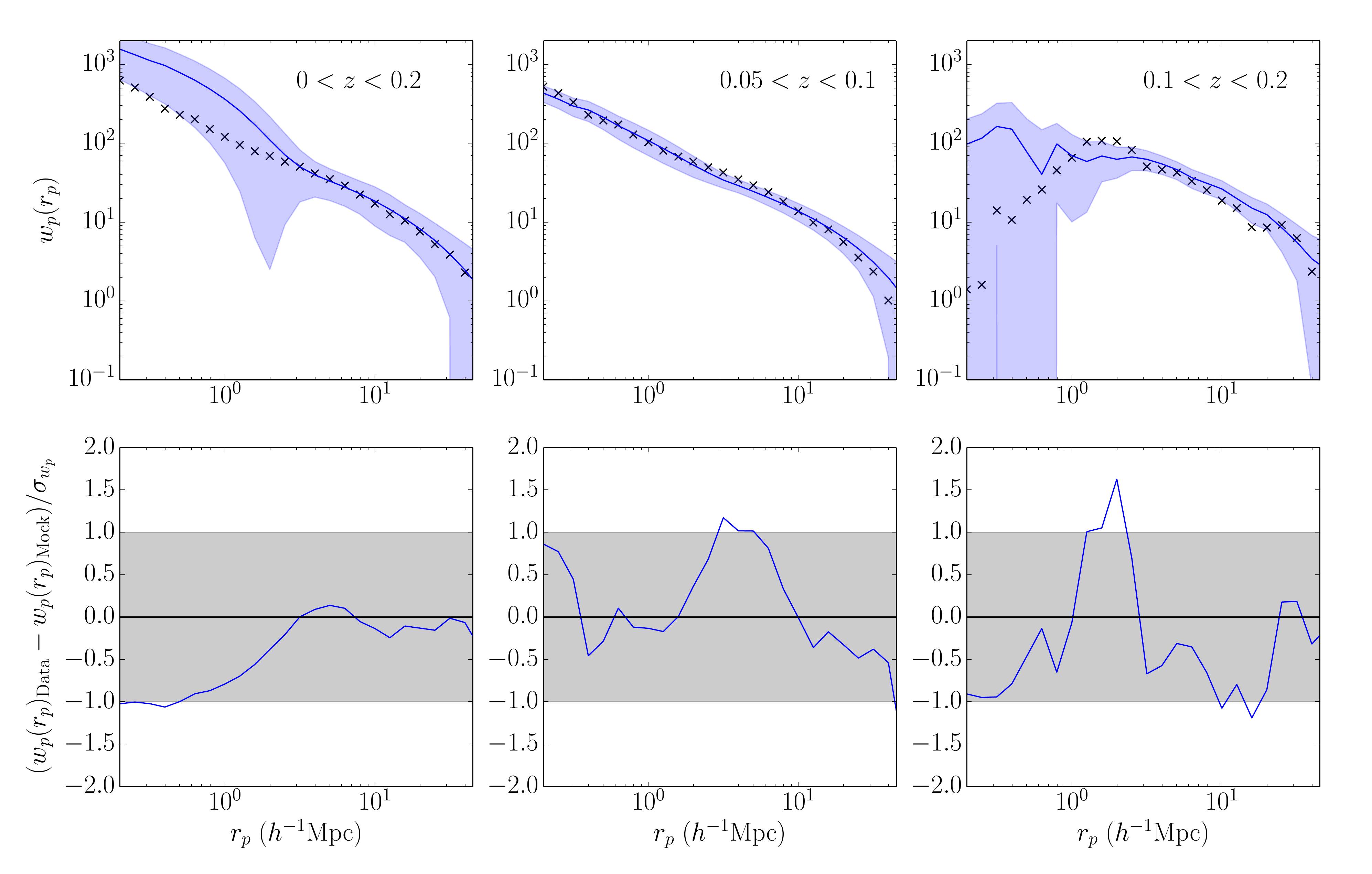}
    \caption{Comparison between the projected correlation function of the data realisation and the mean of the mock catalogues where the blue line shows the COLA mock and the black points the data. There is good agreement at $z > 0.05$, the mocks show a larger amplitude in the clustering in the 1-halo component because of the low mass resolution. The shaded grey regions correspond to the standard deviation.}
    \label{fig:figure3a}
\end{figure*}

\begin{figure}
	\includegraphics[width=\columnwidth]{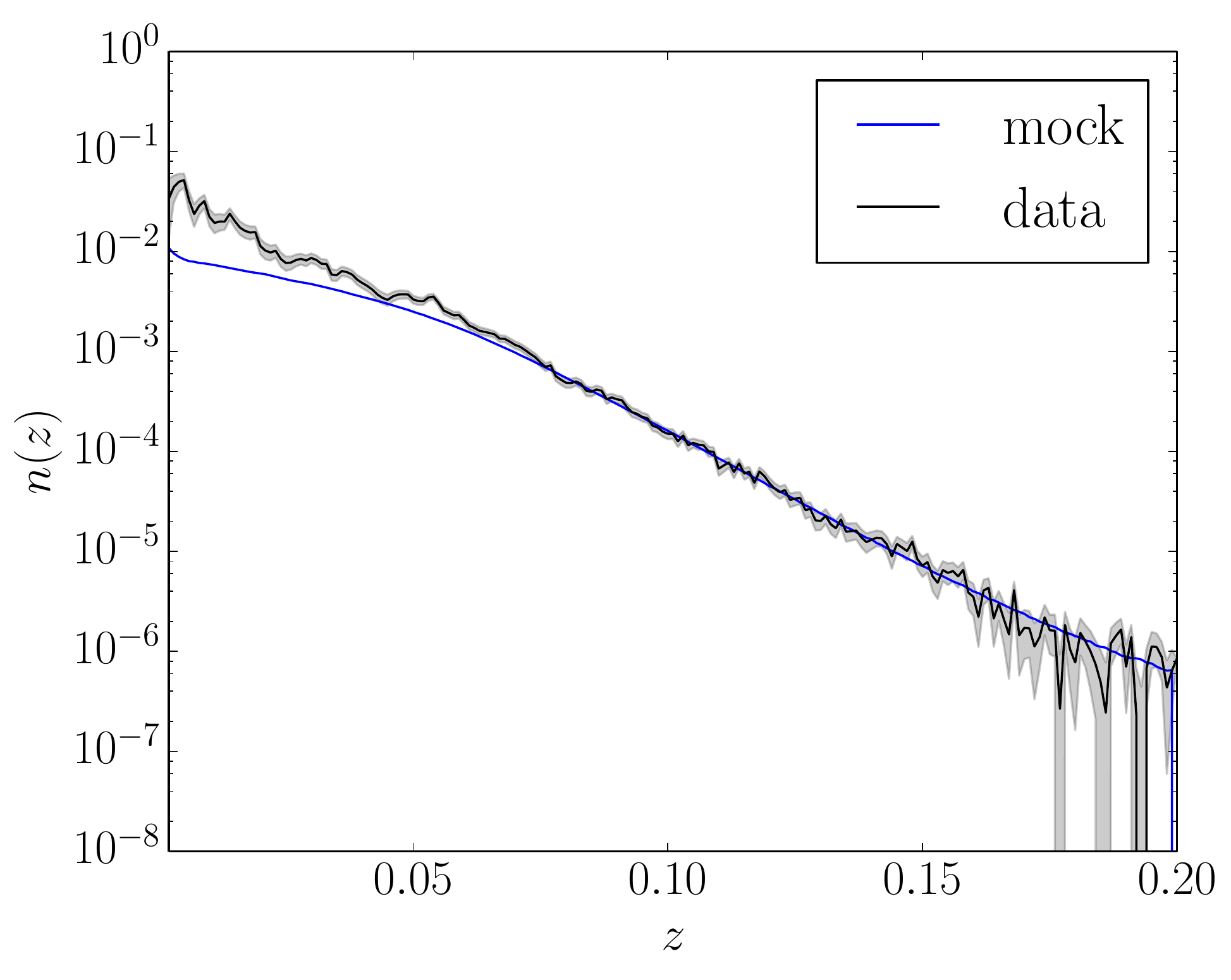}
    \caption{The redshift distribution of the data realisation and the mean of the mock catalogues. Consistency is seen in the redshift range $z > 0.05$, while at $z < 0.05$ the mock shows a lower density to the resolution limit of the COLA mocks.}
    \label{fig:figure3b}
\end{figure}

The redshift dependence of the HOD model is parameterised through a polynomial form on $M_{\mathrm{min}}$ and $M_{1}$,
\begin{equation}
\log_{10}(M_{\mathrm{min}}(z)) = a + bx +  cx^2
           + dx^3 + ex^4
\label{eq:equ6}
\end{equation}
where $x = z - 0.05$. We generate mocks for a grid of HOD parameters ($\log_{10}\left(M_{1}/M_{\mathrm{min}}\right)$, $\sigma_{\log M}$, $\alpha$) and measure the projected correlation function and number density to fit against the data.
We find the best fitting HOD parameters of $\alpha = 1.50 \pm 0.05$, $\sigma_{\log M} = 0.50 \pm 0.12$ and $\log_{10}\left(M_{1}/M_{\mathrm{min}}\right)=1.50 \pm 0.08$ where the polynomial terms are $a = 12.0448$, $b = 22.8194$, $c = 110.4364$, $d = -1435.6529$ and $e = 3679.3770$. The match between the best fit mock and the data are shown in Figure~\ref{fig:figure3a} and Figure~\ref{fig:figure3b}. The mocks are a good fit at $z > 0.05$, but at a lower redshift this breaks down due to the small scale mass resolution issue in the COLA simulation mentioned earlier.

We assign a central galaxy to a halo according to the probability
$<N_{C}(M)>$, and draw a random number of satellites from a Poisson
distribution with mean $<N_{S}(M)>$ if the halo hosts the central galaxy.
The position and the velocity of the central galaxies are those of the
mean of the halo particles (center of mass). For satellite galaxies, we assume an isotropic NFW profile \citep{1996ApJ...462..563N} with a concentration parameter \citep{2001MNRAS.321..559B}. We randomly assign the distance from the halo centre based on
the mass distribution and a Gaussian random velocity based on the
radius-dependent velocity-dispersion profile in addition to the halo
velocity \citep{2004MNRAS.352.1302V,2013A&A...557A..54D}.

\section{Overview of Clustering Measurement}
\label{sec:over}

We choose to measure the BAO signal using the correlation function to facilitate comparison with \citep{2011MNRAS.416.3017B}. In this section we briefly explain how we measure the correlation function and covariance matrix. We will also introduce the density field reconstruction technique which we apply to the dataset.

\subsection{Two-Point Correlation Function}
\label{sec:corrfn}

We measure the two-point correlation statistic using the \cite{1993ApJ...412...64L} estimator,
\begin{equation}
    \xi(s) = 1 + \frac{DD(s)}{RR(s)}\left(\frac{n_{r}}{n_{d}}\right)^{2} - 2\frac{DR(s)}{RR(s)}\left(\frac{n_{r}}{n_{d}}\right),
    \label{eq:equ7}
\end{equation}
\noindent where $DD(s)$, $DR(s)$ and $RR(s)$ are the pair counts of data-data, data-random and random-random pairs split in co-moving space. The random catalogues used during this analysis were generated using the selection function measured from the data catalogue. This Monte-Carlo samples the sky coverage (as given in Figure~\ref{fig:figure1a}) and redshift distribution using 30 times as many random points as galaxies, to ensure that the Poisson noise contribution from the randoms is significantly smaller than that from the galaxies. The normalisation of pair counts is performed with $n_{d}$ and $n_{r}$ which are the weighted number of objects for data and random catalogues respectively.

The data-data pair counts have been assigned a fibre collision weight for each galaxy pair. The design of the 6dF instrument only allowed fibres to be placed $> 5.7 ~\mathrm{arcmin}$ apart resulting in lost pairs on small angle separation. By allowing for multiple passes in the targeting plan, this effect of lost pairs is minimised for $70\%$ of the survey area. The remaining $30\%$ requires an up-weighting of pair counts to ensure we do not bias the clustering statistics. This angular weight $w_{f}(\theta_{12})$ is calculated as a ratio between expected and observed angular correlation functions \citep{2004MNRAS.355..747J, 2013MNRAS.429.3604B} where $\theta_{12}$ is the angle between galaxies in the pair. This approximation works when radial clustering is matched between observed and unobserved pairs. On small scales this is not true and we should instead use a scheme such as that of \cite{2017MNRAS.472.1106B}. As we are only interested in the BAO scale, this is not necessary for this analysis.

\subsection{Covariance Matrix}
\label{sec:covmat}

A robust covariance matrix was generated from 600 mock realisations, built on COLA-based simulations and populated through a HOD parameterisation. Further details of the mock production are given in Section~\ref{sec:mock}, along with the implementation of the HOD model.

The covariance matrix is given by,
\begin{equation}
    \textbf{C}_{ij} = \sum\limits_{n=1}^{N}\frac{(\xi_{n}(s_{i})-\overline{\xi}(s_{i}))(\xi_{n}(s_{j})-\overline{\xi}(s_{j}))}{N-1},
    \label{eq:equ8}
\end{equation}
in which the summation runs over $N$ mock realisations. $\xi_{n}(s_{i})$ is the $i^{\textrm{th}}$ separation bin of the $n^{\textrm{th}}$ mock correlation function and $\overline{\xi}(s_{i})$ is the average in this bin.
\noindent The uncertainty on the correlation functions shown in Figures \ref{fig:figure5}, \ref{fig:fig6} and \ref{fig:figMoM} are from the standard deviation on the diagonal $\sqrt{\textbf{C}_{ii}}$, while during the fitting of models the full covariance matrix is used.

\begin{figure}
	\includegraphics[width=\columnwidth]{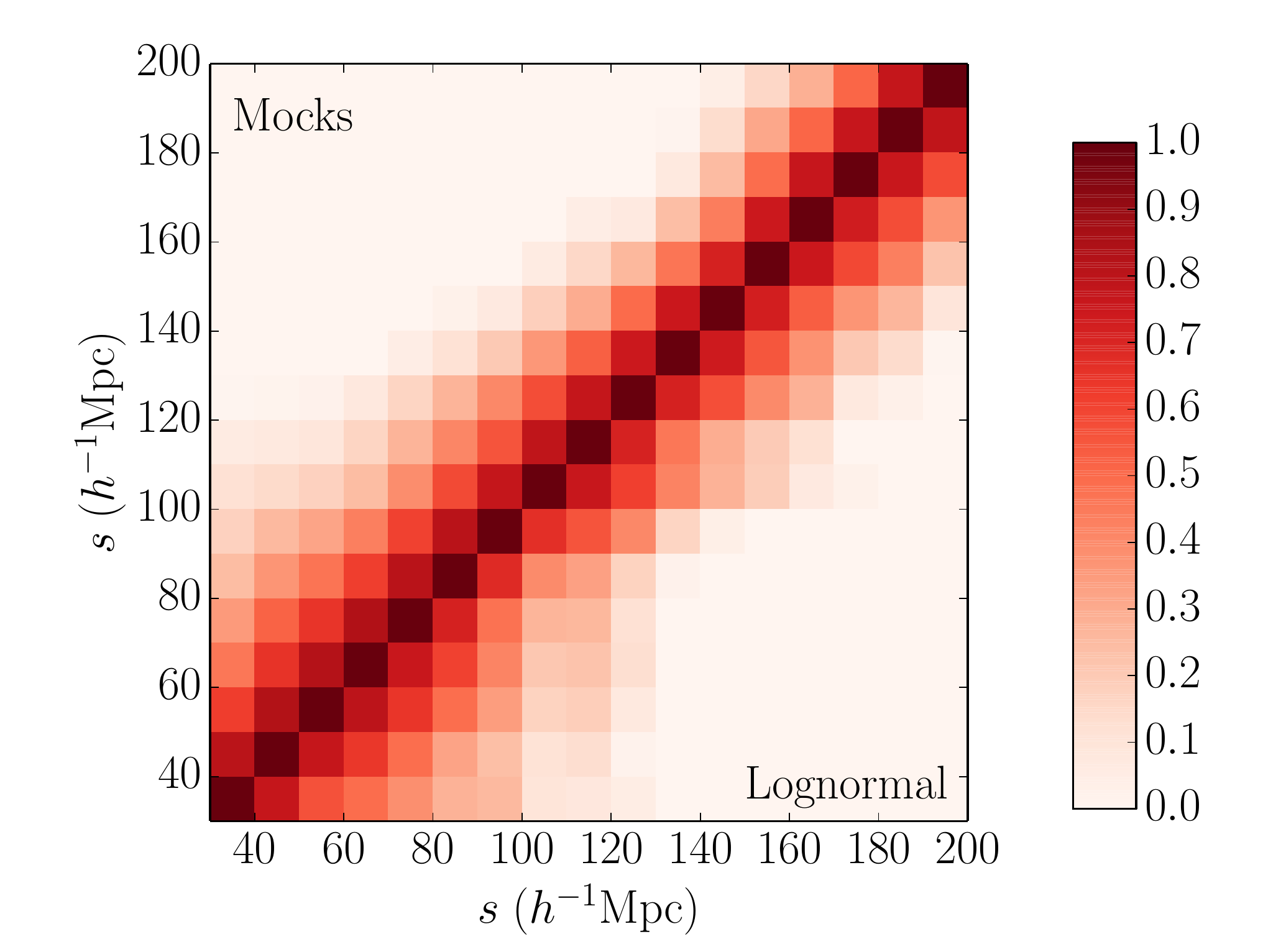}
    \caption{Comparison of the COLA mock and lognormal based correlation matrices. The correlation in the mock covariance matrix has been reduced in comparison to the lognormal based matrix.}
    \label{fig:figure4}
\end{figure}

\begin{figure}
	\includegraphics[width=\columnwidth]{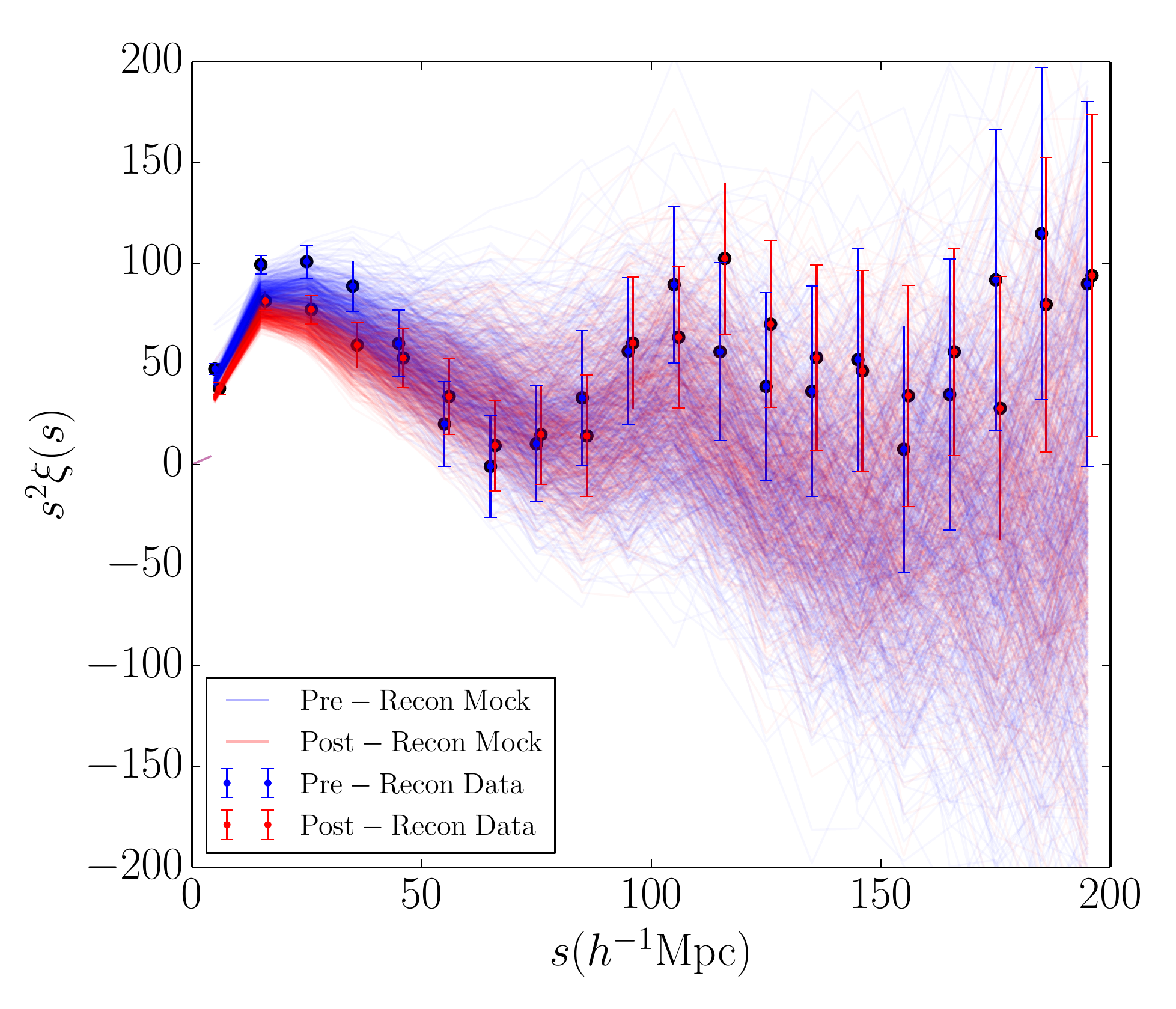}
    \caption{The distribution of all 600 mock correlation functions, both pre- (blue) and post-reconstruction (red). The data points show the 6dFGS correlation function with errors from the diagonal of the constructed COLA mock covariance matrix. The post-reconstruction data points have been displaced by $+1\,h^{-1}\mathrm{Mpc}$ for clarity.}
    \label{fig:figure5}
\end{figure}

The covariance matrix, calculation without reconstruction is compared to that previously used based on lognormal catalogues \citep{2011MNRAS.416.3017B} in Figure~\ref{fig:figure4}. This comparison shows that the correlation between neighbouring bins has decreased using the COLA mocks. The correlation functions of the mock catalogues are compared to that from the data in Figure~\ref{fig:figure5} showing good agreement. These mock catalogues have been used in other recent work on the 6dFGS \citep{2017PhRvD..95h3502A, 2018arXiv180104969B}.

\subsection{FFT-based Density Field Reconstruction}
\label{sec:recon}

The formalism used for density field reconstruction follows from \citep{2014MNRAS.445.3152B, 2015MNRAS.453..456B}. In a Lagrangian framework the Eulerian position of a particle is given by,
\begin{equation}
\textbf{\textrm{x}}(\textbf{\textrm{q}}, t) = \textbf{\textrm{q}} + \boldsymbol{\Psi}(\textbf{\textrm{q}}, t),
\label{eq:equ9}
\end{equation}
where the \textbf{$\textrm{q}$} is the Lagrangian position and $\boldsymbol{\Psi}$ is the displacement vector field. Implementing first order Lagrangian Perturbation Theory (LPT) the standard Zel'dovich approximation \citep{1970A&A.....5...84Z} can be obtained,
\begin{equation}
\boldsymbol{\Psi}_{(1)}(\textbf{\textrm{k}})=-\frac{i\textbf{\textrm{k}}}{k^{2}}\delta_{(1)}(\textbf{\textrm{k}}),
\label{eq:equ10}
\end{equation}
which relates the Fourier transform of the overdensity field to the displacement field in $\textbf{\textrm{k}}$-space. To linear order galaxies trace the matter density field as $\delta_{g}=b\delta_{m}$ where $b$ is the bias. Because of the redshift space distortions (RSDs), to obtain the displacement field $\Psi$ we actually have to solve the differential equation,
\begin{equation}
\nabla\cdot\boldsymbol{\Psi}+\frac{f}{b}\nabla\cdot(\boldsymbol{\Psi}\cdot\boldsymbol{\hat{r}})\boldsymbol{\hat{r}} = -\frac{\delta_{g}}{b},
\label{eq:equ11}
\end{equation}
On linear scales RSD enhances the clustering along the line-of-sight, dependent on the amplitude of $f = d\ln(D(a))/d\ln(a)$ the growth rate, $D(a)$ the growth function, $a$ the scale factor and $\sigma_{8}$ describes the amplitude of the linear power spectrum on scales of $8\,h^{-1}$Mpc.

Equation~\ref{eq:equ11} can be solved as in \cite{2012MNRAS.427.2132P} using a finite difference approximation to compute the gradients. This sets up a grid in configuration space through which the potential can be described as a linear system of equations. This methodology was chosen because although $\boldsymbol{\Psi}$ is irrotational, the term $(\boldsymbol{\Psi}\cdot\boldsymbol{\hat{r}})\boldsymbol{\hat{r}}$ is not, hence you cannot locate the solution directly with Fourier methods. However \cite{2015MNRAS.453..456B} showed that by making the approximation that $(\boldsymbol{\Psi}\cdot\boldsymbol{\hat{r}})\boldsymbol{\hat{r}}$ can be irrotational and iterating after correcting, one can efficiently obtain the correct solution using FFTs (with $\textrm{IFFT}$ referring to the inverse fast fourier transform) with $\beta = f/b$,
\begin{equation}
\boldsymbol{\Psi}=\textrm{IFFT}\left[-\frac{i\textbf{\textrm{k}}\delta(k)}{k^{2}b}\right]-\frac{\beta}{1+\beta}\left(\textrm{IFFT}\left[-\frac{i\textbf{\textrm{k}}\delta(k)}{k^{2}b}\right]\cdot\hat{\textbf{r}}\right)\hat{\textbf{r}},
\label{eq:equ12}
\end{equation}
The displacement field calculated from this form of the algorithm has been shown to agree with the finite difference approach and causes negligible differences between post-reconstruction 2-point statistics \cite{2014MNRAS.445.3152B}.

To also remove RSD we modify the displacement vector as $\Psi^{\mathrm{final}}=\Psi+\Psi_{\mathrm{RSD}}$ \citep{1987MNRAS.227....1K, 2012MNRAS.427.2132P} where,
\begin{equation}
\Psi_{\mathrm{RSD}} = -f(\mathbf{\Psi}\cdot\mathbf{\hat{r}})\mathbf{\hat{r}},
\label{eq:equ13}
\end{equation}
using the already calculated displacement field along the line-of-sight. This retrieval of the real-space post-reconstruction density field results in the reduction of amplitude in the correlation function (Figure~\ref{fig:figure5}).

\section{Correlation Function Modelling}
\label{sec:model}

In \citet{2011MNRAS.416.3017B} a model for the pre-reconstruction correlation function was constructed from the formalism of \citet{2008PhRvD..77b3533C}. Here we use a model that marginalises over the shape of the correlation function through a number of nuisance parameters \citep{2014MNRAS.441...24A}. This freedom to marginalise over the broadband shape is important when modelling the post-reconstruction correlation function because we do not have a physically motivated model. 

The fit of the model $\textbf{m}(s)$ to the data $\textbf{d}(s)$ is determined through a minimum-$\chi^{2}$ fitting, $\chi^{2} = D^{T}\textbf{C}^{-1}D$, where $D = \textbf{d}(s) - \textbf{m}(s)$. In our standard analysis we use a fitting range of $30\,h^{-1}\mathrm{Mpc} < s < 200\,h^{-1}\mathrm{Mpc}$ over 17 bins. The inverse covariance matrix used was built from the set of 600 mocks correcting for statistical bias in the standard way \citep{2007A&A...464..399H}.

\subsection{Template Formalism}
The fitting of the correlation function relies primarily on a template model with damped BAO given in \cite{2007ApJ...664..660E}. The model power spectrum that is used in this template contains a linear model $P_{\mathrm{lin}}(k)$ from CAMB \citep{Lewis:1999bs} and a no-wiggle power spectrum $P_{\mathrm{nw}}(k)$ \citep{1998ApJ...496..605E},
\begin{equation}
    P^{\textrm{mod}}(k) = P_{\textrm{nw}}(k)\left[1+\left(\frac{P_{\textrm{lin}}(k)}{P_{\textrm{nw}}(k)}-1\right)e^{-\frac{1}{2}k^{2}\Sigma^{2}_{\textrm{nl}}}\right].
    \label{eq:equ14}
\end{equation}
\noindent The no-wiggle model has had the BAO feature removed whilst retaining the overall shape. The template $\xi^{\textrm{mod}}(s)$ is then obtained from this through a Hankel transform and used with 5 nuisance parameters ($B_{\xi}$, $a_{1}$, $a_{2}$, $a_{3}$, $\Sigma_{\mathrm{nl}}$) which allow marginalisation of the BAO damping (the scale of which is set by $\Sigma_{\mathrm{nl}}$) and broadband shape of the correlation function,
\begin{equation}
    \xi^{\textrm{fit}}(s) = B^{2}\xi^{\textrm{mod}}(\alpha s)+\frac{a_{1}}{s^{2}} + \frac{a_{2}}{s} + a_{3}.
    \label{eq:equ15}
\end{equation}
$\alpha$ is defined as,
\begin{equation}
    \alpha = \frac{D_{V}(z_{\textrm{eff}})r_{s}^{\textrm{fid}}}{D_{V}^{\textrm{fid}}(z_{\textrm{eff}})r_{s}},
    \label{eq:equ16}
\end{equation}
\noindent and allows one to freely scale the BAO feature in our model. A best fit value of $\alpha < 1$ corresponds to a larger BAO scale compared to the fiducial cosmology used to construct our template and $\alpha > 1$ indicates a smaller BAO scale. $D_{V}$ is the volume averaged distance,
\begin{equation}
    D_{V} = \left[cz(1+z)^{2}D_{A}^{2}(z)H^{-1}(z)\right]^{1/3}.
    \label{eq:equ17}
\end{equation}

\subsection{Pre-Reconstruction}
To provide constraints on the pre-reconstruction BAO peak the above template was utilised through a Likelihood based routine similar to that outlined in \cite{2015MNRAS.449..835R}. The model has 6 free parameters ($B$, $\alpha$, $a_{1}$, $a_{2}$, $a_{3}$, $\Sigma_{\mathrm{nl}}$) and we assume a Gaussian prior on the non-linear damping scale $\Sigma_{\mathrm{nl}}=10.3\pm2.0\,h^{-1}\mathrm{Mpc}$ as measured from the mean of the mocks (discussed in Section~\ref{sec:tests}). The $1\sigma$ and $2\sigma$ uncertainties are defined at the $\Delta\chi^{2}=1$ and $4$ levels respectively after marginalisation over the nuisance parameters.

\noindent The best fit correlation function is compared to the data in Figure~\ref{fig:fig6}. We also show the $\chi^{2}$ distribution. The marginalised constraint on the shift parameter $\alpha$ pre-reconstruction is $\alpha = 1.018\pm 0.067$ with $\chi^{2}/\nu = 8.86/11 = 0.81$, which is consistent with \citet{2011MNRAS.416.3017B} who used a WMAP7 fiducial cosmology and found $\alpha = 1.036 \pm 0.062$ (the $\alpha$ fit from our work translates to $\alpha_{\mathrm{WMAP7}} = \alpha_{\mathrm{Planck}}[(D_{V}^{\mathrm{Planck}}/r_{s}^{\mathrm{Planck}})/(D_{V}^{\mathrm{WMAP7}}/r_{s}^{\mathrm{WMAP7}})] = 1.047\pm0.069$ for comparison in the WMAP7 fiducial model). The difference in uncertainty is due to the revised covariance matrix (see Section \ref{sec:covmat} for a comparison).

\subsection{Post-Reconstruction}
\par We apply density field reconstruction to both the data and COLA mock catalogues. In doing so a galaxy bias of $b = 1.82$ \citep{2011MNRAS.416.3017B} and a growth rate of $f(z_{\mathrm{eff}}=0.097)=0.579$ were assumed. Our results are independent of this choice as shown later in Section \ref{sec:robust}. When calculating the displacement field, the overdensity field was smoothed using a Gaussian smoothing kernel, $S(k)=e^{{-(kR)^2}/2}$, with the smoothing scale of $R = 15\,h^{-1}\mathrm{Mpc}$. Using the calculated displacement field as a proxy for the non-linear evolution, the catalogues were shifted to move the field back into the psuedo-linear regime. The reconstruction algorithm gave a mean galaxy shift of $\bar{s} = 5.87\,h^{-1}\mathrm{Mpc}$.
\par The correlation function post-reconstruction has an estimator that includes both a shifted random $S$ and independent unshifted random catalogues $R$ \citep{2012MNRAS.427.2132P},

\begin{equation}
    \xi(s) = \frac{SS(s)}{RR(s)}\left(\frac{n_{r}}{n_{s}}\right)^{2} + \frac{DD(s)}{RR(s)}\left(\frac{n_{r}}{n_{d}}\right)^{2} - 2\frac{DS(s)}{RR(s)}\left(\frac{n_{r}^{2}}{n_{d}n_{s}}\right).
    \label{eq:equ18}
\end{equation}

\begin{figure*}
	\includegraphics[width=2\columnwidth]{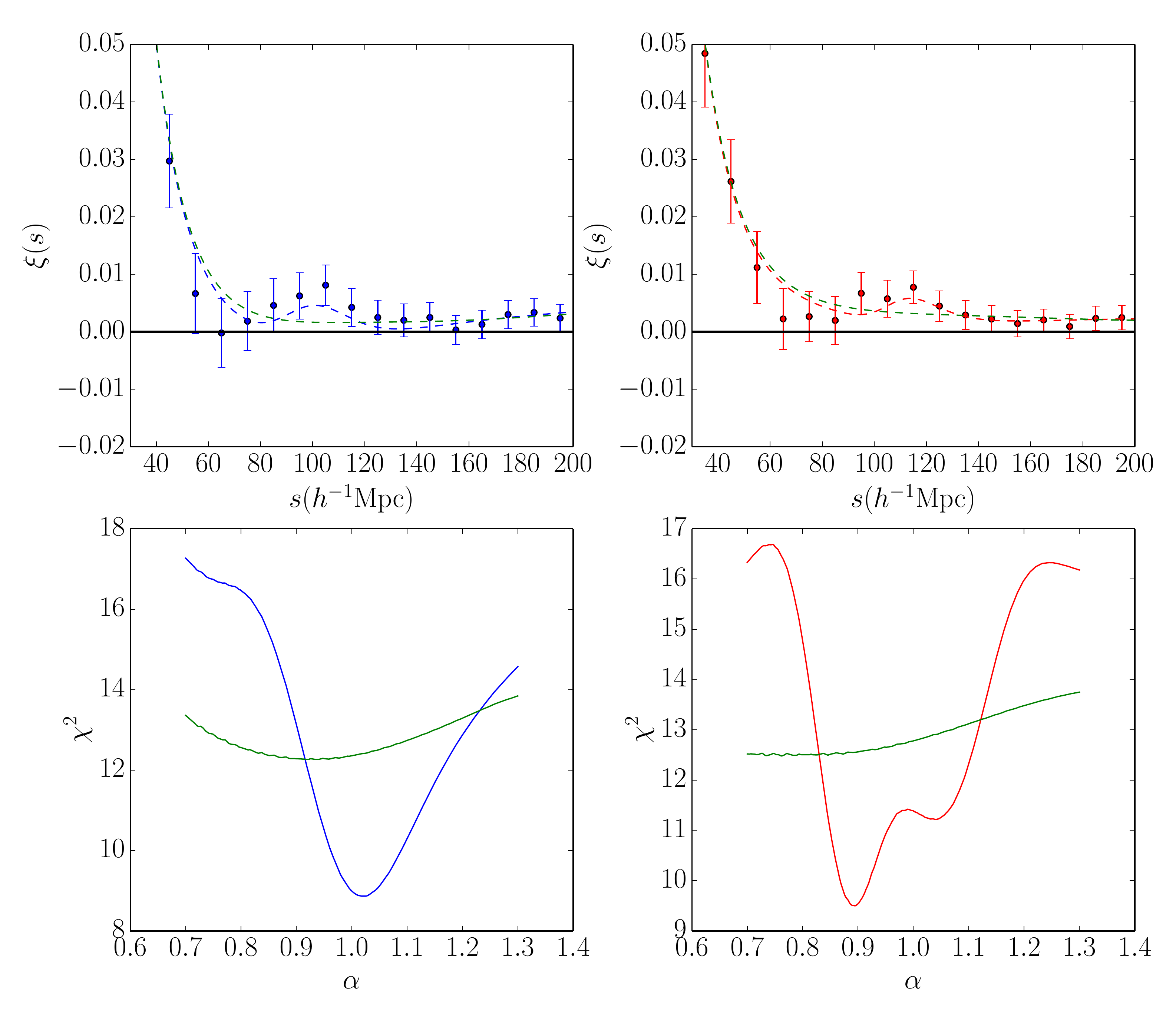}
    \caption{Left: In the upper plot the pre-reconstruction correlation function is given with the best fit model, corresponding to $\alpha = 1.018\pm 0.066$. This shows consistency with the results from \citet{2011MNRAS.416.3017B} in which they used a lognormal covariance matrix. The plot below provides the marginalised $\chi^{2}$ distribution for $\alpha$. Right: In the upper plot the post-reconstruction correlation function is given with the best fit model, corresponding to $\alpha = 0.895\pm0.041$. The plot below provides the marginalised $\chi^{2}$ distribution for $\alpha$. There is a strong bi-modality in this likelihood surface, which translates to strongly non-gaussian errors at the 2$\sigma$ ($\Delta\chi^{2}=4$) level. This means that although the best fit for the 6dFGS post-reconstruction constraint was found at $\alpha < 0.9$, the likelihood also gives weight to $\alpha\sim1.04$ when combining with MGS measurements (Section~\ref{sec:joint}).}
    \label{fig:fig6}
\end{figure*}

As in the pre-reconstruction case, we use the smoothing parameter fit from the mean of the mocks with a prior $\Sigma_{nl}=4.8\pm 2.0h^{-1}\mathrm{Mpc}$. The marginalised constraint on the shift parameter $\alpha$ post-reconstruction is $\alpha = 0.895 \pm 0.042 (\pm ^{0.235}_{0.079})$ with $\chi^{2}/\nu = 9.49/11=0.86$ (the error in the bracket showing the non-Gaussianity at the $98\%$ confidence level). The model, data and $\chi^{2}$ distribution are displayed in Figure~\ref{fig:fig6}. The $\alpha$ post-reconstruction found has a non-Gaussian likelihood beyond the $1\sigma$ region which means that although the value is displaced from the average, it is still consistent with the mock catalogues (discussed further in Section \ref{sec:tests}).

\section{Tests on the mock catalogues}
\label{sec:tests}
\subsection{Fitting the mean of the mocks}
\begin{figure}
	\includegraphics[width=\columnwidth]{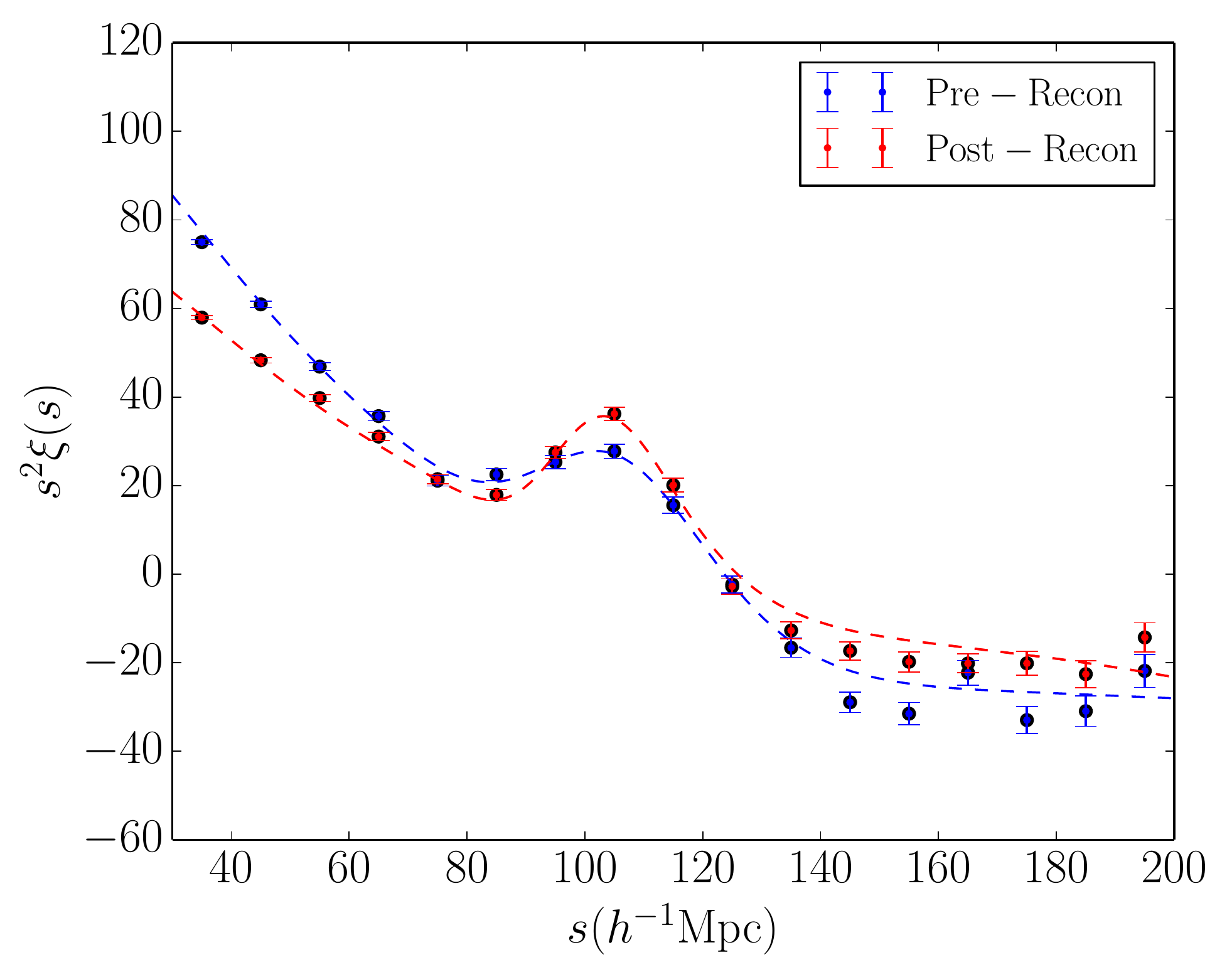}
    \caption{The best fit to the pre/post-reconstruction mean of the mock correlation functions. The improvement factor is $I\sim1.86$ and the best fit non-linear damping scales of $\Sigma_{\mathrm{nl}}=10.3\,h^{-1}\mathrm{Mpc}$ and $\Sigma_{\mathrm{nl}}=4.8\,h^{-1}\mathrm{Mpc}$, respectively.}
    \label{fig:figMoM}
\end{figure}

We fit both the pre/post reconstruction average correlation functions from the mock population. This fit makes use of the covariance matrices, which have been rescaled by the number of realisations $N$, $\mathbf{C}_{\mathrm{mean}} = \mathbf{C}_{\mathrm{one}}/N$. The best fit model is shown in Figure~\ref{fig:figMoM} with $\alpha_{\mathrm{pre}}=0.999 \pm 0.0065$ and $\alpha_{\mathrm{post}}=0.997\pm0.0035$ giving an improvement factor of $I = \sigma_{\alpha, \mathrm{pre}}/\sigma_{\alpha, \mathrm{post}} \sim 1.86$.

\subsection{Comparison between data and fits to individual mocks}
\begin{figure*}
	\includegraphics[width=2\columnwidth]{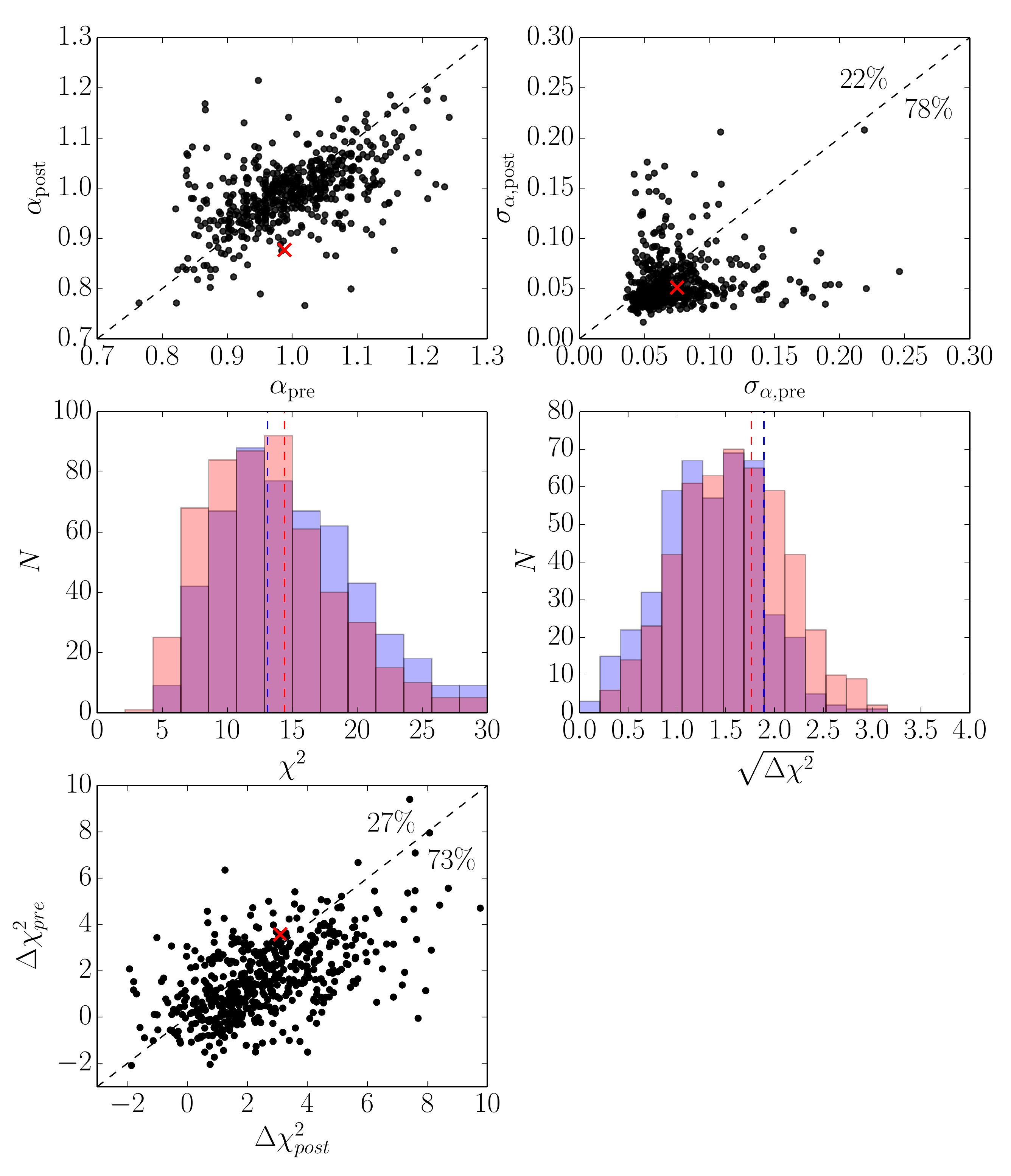}
    \caption{The distribution of (upper left) $\alpha$, (upper right) $\sigma_{\alpha}$, (middle left) $\chi^{2}$, (middle right) $\sqrt{\Delta\chi^{2}}$ (significance of detetction), (lower left) $\Delta\chi^{2}$ comparison pre/post-reconstruction for the mock sample that have been cut for outliers and to realisations that have $1\sigma$ contours fully within the prior range ($0.7 < \alpha < 1.3$). The data (red crosses and dashed lines (blue/red:pre/post) respectively) is within the range of the mock catalogues, that show improvement following density field reconstruction.}
    \label{fig:fig10}
\end{figure*}

To make comparisons between the mock population and the data, each of the 600 realisations used to produce the covariance matrices were individually fit. $\sim 30\%$ of the mock catalogues in the population did not have a well constrained measurement of $\alpha$, a similar fraction to that found for the SDSS MGS analysis \citep{2015MNRAS.449..835R}. To ensure that this comparison was performed only on the mocks that have a relevant detection of the BAO feature, we select a subsample having $1\sigma$ contours ($\Delta\chi^{2}=1$) within the prior region $0.7 < \alpha < 1.3$ (both pre- and post-reconstruction). This cut reduces the population to $70\%$ of its original number. Comparisons of the distributions of best-fitting $\chi^{2}$, the value of $\alpha$ pre/post-reconstruction and $\sigma_{\alpha}$ the $1\sigma$ error bound (Fig.~\ref{fig:fig10}) show that the data realisation is within the locus of measurements from this subsample of the mocks. 

The distribution of errors on $\alpha$, shows that for our mocks we see 80\% of these mocks have $\sigma_{\alpha, \mathrm{pre}}/\sigma_{\alpha, \mathrm{post}} > 1$, as would be expected after applying density field reconstruction. Our detection in the data of the BAO peak using our model, which marginalises over broadband shape, is at the $\sim 1.9\sigma$ level pre-reconstruction and $\sim 1.75\sigma$ post-reconstruction. In the subsample of the mocks pre-reconstruction $18\%$ have a detection higher than this and post-reconstruction this increases to $52\%$. The mean detection level pre/post-reconstruction for the mocks is $1.5\sigma$ and $1.8\sigma$ respectively. This increase in the number of high significance detections in the mocks shows the expected trend from density field reconstruction to, on average, enhance the significance of detection. The lower left plot in Figure~\ref{fig:fig10} shows however that in $28\%$ of cases, in the mock sample, reconstruction lowers the significance of detection. The reduction of significance seen in the data realisation is therefore likely a case of the data being one of these ``unlucky" samples. When comparing the pre-reconstruction significance detection to \cite{2011MNRAS.416.3017B} we find a lower value, this is due to a number of conservative alterations. These include (1) a change in the fitting model to include polynomial terms that marginalise over the shape giving more freedom, (2) a change of fitting range from $10\,h^{-1}\mathrm{Mpc}<s<200\,h^{-1}\mathrm{Mpc}$ to $30\,h^{-1}\mathrm{Mpc}<s<200\,h^{-1}\mathrm{Mpc}$ and (3) the new robust covariance matrix made from the COLA-based mocks rather than log-normal realisations (but which has slightly larger covariance amplitude).

\subsection{Robustness Tests}
\label{sec:robust}
To investigate the robustness of the post-reconstruction $\alpha$ detection and the uncertainty on this parameter, tests are run where parameters are systematically varied in both the fitting procedure and density field reconstruction.

During the post-reconstruction analysis, our default procedure uses 17 bins between $30\,h^{-1}\mathrm{Mpc} < s < 200\,h^{-1}\mathrm{Mpc}$ and takes the non-linear damping as that from the mean of the mocks with a small prior. To test the robustness of the procedure, the scale over which the fitting occurs and also the damping scale are varied. The damping scale was fixed both at the value for mean of the mocks and also held at $\Sigma_{nl} = 0\,h^{-1}\mathrm{Mpc}$ (the value that given complete freedom this parameter tends towards). In testing the robustness against binning, the post-reconstruction analysis was conducted with correlation function binning $\Delta s = 8\,h^{-1}\mathrm{Mpc}, 12\,h^{-1}\mathrm{Mpc}$ and $10\,h^{-1}\mathrm{Mpc}$ displaced by $5\,h^{-1}\mathrm{Mpc}$ in comparison to the standard pipeline.

When testing the robustness of the post-reconstruction result the input survey linear bias is varied from $b=1.82$ by $\pm 0.15$ and also the smoothing kernel used on the density field is varied from $15h^{-1}\mathrm{Mpc}$ by $\pm 5h^{-1}\mathrm{Mpc}$. It should be noted that the covariance matrix used during these reconstruction input parameter tests is still derived from the default reconstruction. Hence this mostly tests the robustness of $\alpha$ and does not test $\sigma_{\alpha}$. Test results are collated in Table.~\ref{tab:2} and described below:

\begin{enumerate}
\item Having a best-fit $\alpha \sim 0.9$ is robust for all test cases. However the uncertainty of $\alpha$ depends on the scales fitted. Changing the scale used in these three cases varied the uncertainty by $\Delta(\sigma_{\alpha})=0.01$.

\item Given complete freedom to the non-linear damping the scale which minimises $\chi^{2}$ is $0h^{-1}\mathrm{Mpc}$, although this is only weakly preferred in comparison to the value found from the mean of the mocks, $\Delta\chi^{2}=0.08$. However, although it does not make a large difference in $\chi^{2}$ it does impact $\sigma_{\alpha}$ at the $\Delta(\sigma_{\alpha})=0.01$ level. As the use of $\Sigma_{nl} = 0h^{-1}\mathrm{Mpc}$ would likely underestimate the uncertainty present, our default procedure is to place a small prior on the value centred on the mean recovered from the mocks. The value of $\alpha$ though remains robust against the chosen value of the smoothing scale.

\item We have also tested changing the binning of the correlation function and covariance matrix from our default of $10\,h^{-1}\mathrm{Mpc}$ to $8\,h^{-1}\mathrm{Mpc}$, $12\,h^{-1}\mathrm{Mpc}$ and also displacing the bin centres by $5\,h^{-1}\mathrm{Mpc}$ whilst retaining the default width. In all cases the value of $\alpha$ post-reconstruction varies by up to $\sim 4\%$ about the default case and $\sigma_{\alpha}$ is within $\Delta(\sigma_{\alpha})=0.01$, except for the $8\,h^{-1}\mathrm{Mpc}$. In this case the non-Gaussian likelihood has given more weighting to the sub-dominant peak widening the overall distribution at the $1\sigma$ level.

\item When varying the linear bias used during density field reconstruction by $\pm 0.15$ and also changing the smoothing kernel scale, $\alpha$ is found to only change by $1.5\%$ except for in the case of using $R = 20h^{-1}\mathrm{Mpc}$. In this case there is a large shift in $\alpha$ which is driven by the sub-dominant peak in the non-Gaussian likelihood of $\alpha$ becoming dominant.

\end{enumerate}

\begin{table}
 \centering
 \caption{The results of a number of robustness tests applied to the 6dFGS post-reconstruction data. The first set of tests vary the scale over which our template is fit, the second set uses variations of the damping scale fixed at both the exact value located from the mean of the mocks and $0\,h^{-1}\mathrm{Mpc}$ (which is the value preferred if left completely free). Thirdly we vary the binning used in the correlation function when fitting against the model from $10\,h^{-1}\mathrm{Mpc}$ to $8\,h^{-1}\mathrm{Mpc}$, $12\,h^{-1}\mathrm{Mpc}$ and also displacing the bin centres by $5\,h^{-1}\mathrm{Mpc}$. During reconstruction parameter tests we vary the linear galaxy bias $b$ input $\pm 0.15$ and finally the scale of smoothing $R$ in the Gaussian kernel by $\pm 5\,h^{-1}\mathrm{Mpc}$.}
 \label{tab:2}
 \begin{tabular}{cccccc}
  \hline
  Test & $\alpha$ & $\chi^{2}/\nu$\\
  \hline
  Normal Pipeline & $0.895\pm0.042$ & $9.49/11$\\
  \hline
  Fitting Procedure & & \\
  \hline
  (i) & & \\
  $20\,h^{-1}\mathrm{Mpc} < s < 200\,h^{-1}\mathrm{Mpc}$ & $0.902\pm0.041$ & $10.78/12$\\
  $40\,h^{-1}\mathrm{Mpc} < s < 190\,h^{-1}\mathrm{Mpc}$ & $0.895\pm0.032$ & $8.33/9$\\
  $50\,h^{-1}\mathrm{Mpc} < s < 180\,h^{-1}\mathrm{Mpc}$ & $0.895\pm0.035$ & $8.20/7$\\
  \hline
  (ii) & & \\
  $\Sigma_{nl} = 4.5\,h^{-1}\mathrm{Mpc}$ fixed & $0.902\pm0.041$ & $9.50/12$\\
  $\Sigma_{nl} = 0\,h^{-1}\mathrm{Mpc}$ & $0.887\pm0.031$ & $9.42/12$\\
  \hline
  (iii) & & \\
  $\Delta s = 8\,h^{-1}\mathrm{Mpc}$ & $0.913\pm0.112$ & $11.93/15$\\
  $\Delta s = 12\,h^{-1}\mathrm{Mpc}$ & $0.929\pm0.036$ & $5.91/7$\\
  $\Delta s = 10\,h^{-1}\mathrm{Mpc}$ $(+5\,h^{-1}\mathrm{Mpc})$ & $0.865\pm0.050$ & $9.44/11$\\
  \hline
  Reconstruction Parameters & & \\
  \hline
  (iv) & & \\
  $b = 1.67$ & $0.895\pm0.052$ & $9.89/11$\\
  $b = 1.97$ & $0.894\pm0.044$ & $9.893/11$\\
  \hline
  $R = 10\,h^{-1}\mathrm{Mpc}$ & $0.908\pm0.044$ & $8.95/11$\\
  $R = 20\,h^{-1}\mathrm{Mpc}$ & $1.053\pm0.045$ & $9.88/11$\\
 \end{tabular}
\end{table}

\begin{figure}
	\includegraphics[width=\columnwidth]{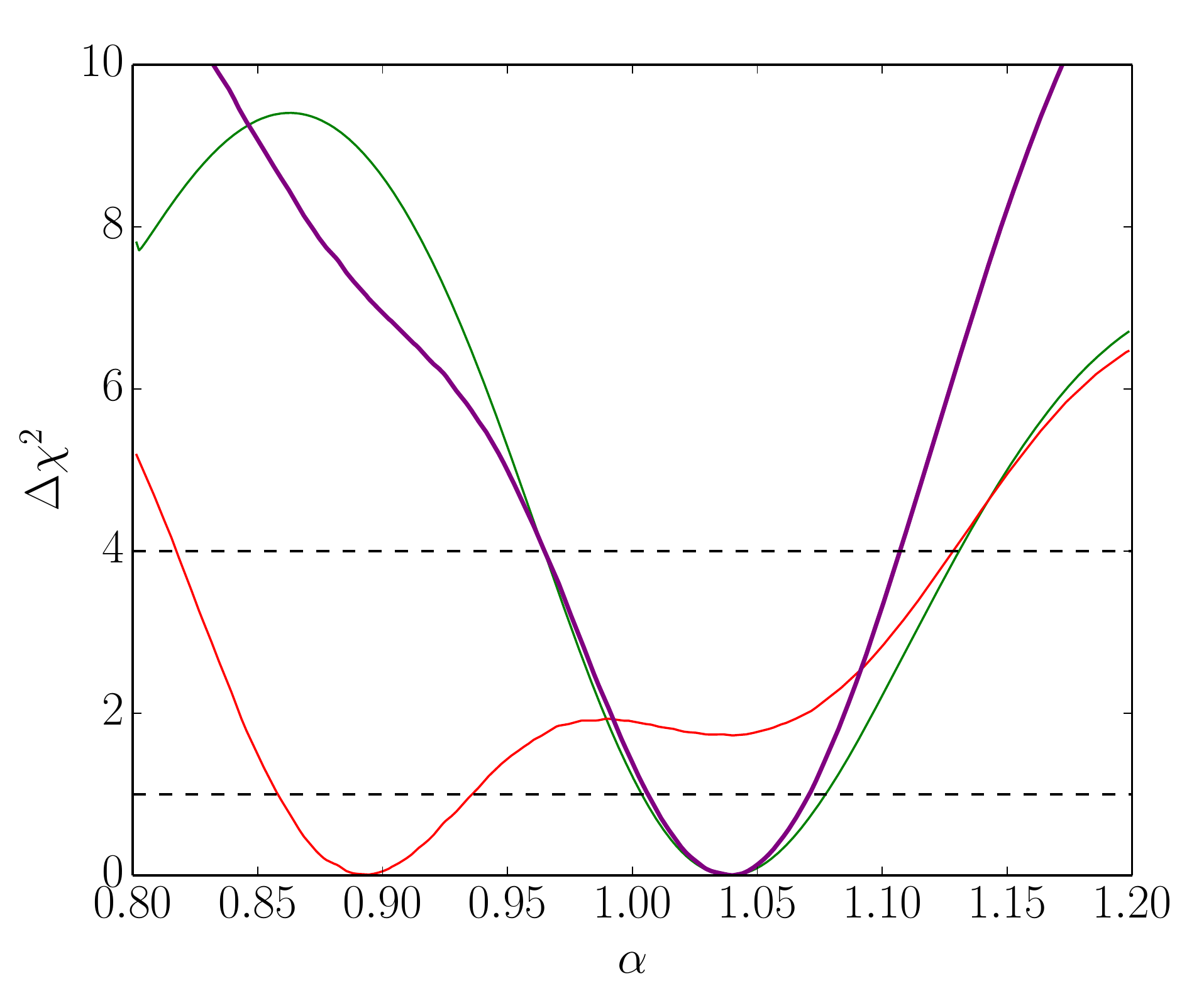}
    \caption{The $\Delta\chi^{2}$ distributions of $\alpha$ post-reconstruction 6dFGS (red), SDSS MGS (green) and joint (purple).}
    \label{fig:figJ}
\end{figure}

\section{Cosmological Interpretation}
\label{sec:cosmo}
In this section we convert our measurements of $\alpha$ to a volume averaged distance measurements and combine with a $\Omega_{m}h^{2}$ prior from Planck 2015 to offer $H_{0}$ and $\Omega_{m}$ cosmological constraints. Combining the joint clustering measurement with $\Omega_{m}h^{2}$ from the CMB calibrates the standard ruler.

\subsection{6dFGS post-reconstruction only}
The best fit constraint post-reconstruction translates to $D_{V}(z_{\mathrm{eff}}=0.097) = \alpha D_{V}^{\mathrm{fid}}(z_{\mathrm{eff}}=0.097)(r_{s}/r_{s}^{\mathrm{fid}}) = 372\pm17(r_{s}/r^{\mathrm{fid}}_{s})\,\mathrm{Mpc}$ with $D_{V}^{\mathrm{fid}}(z_{\mathrm{eff}}=0.097) = 416\,\mathrm{Mpc}$. This result is consistent with our fiducial cosmology at the $2\sigma$ level. Although the measured value of $\alpha$ is reasonably far from 1, the likelihood shows a strong bimodal shape (see Figure~\ref{fig:fig6}).

\begin{figure*}
	\includegraphics[width=1.2\columnwidth]{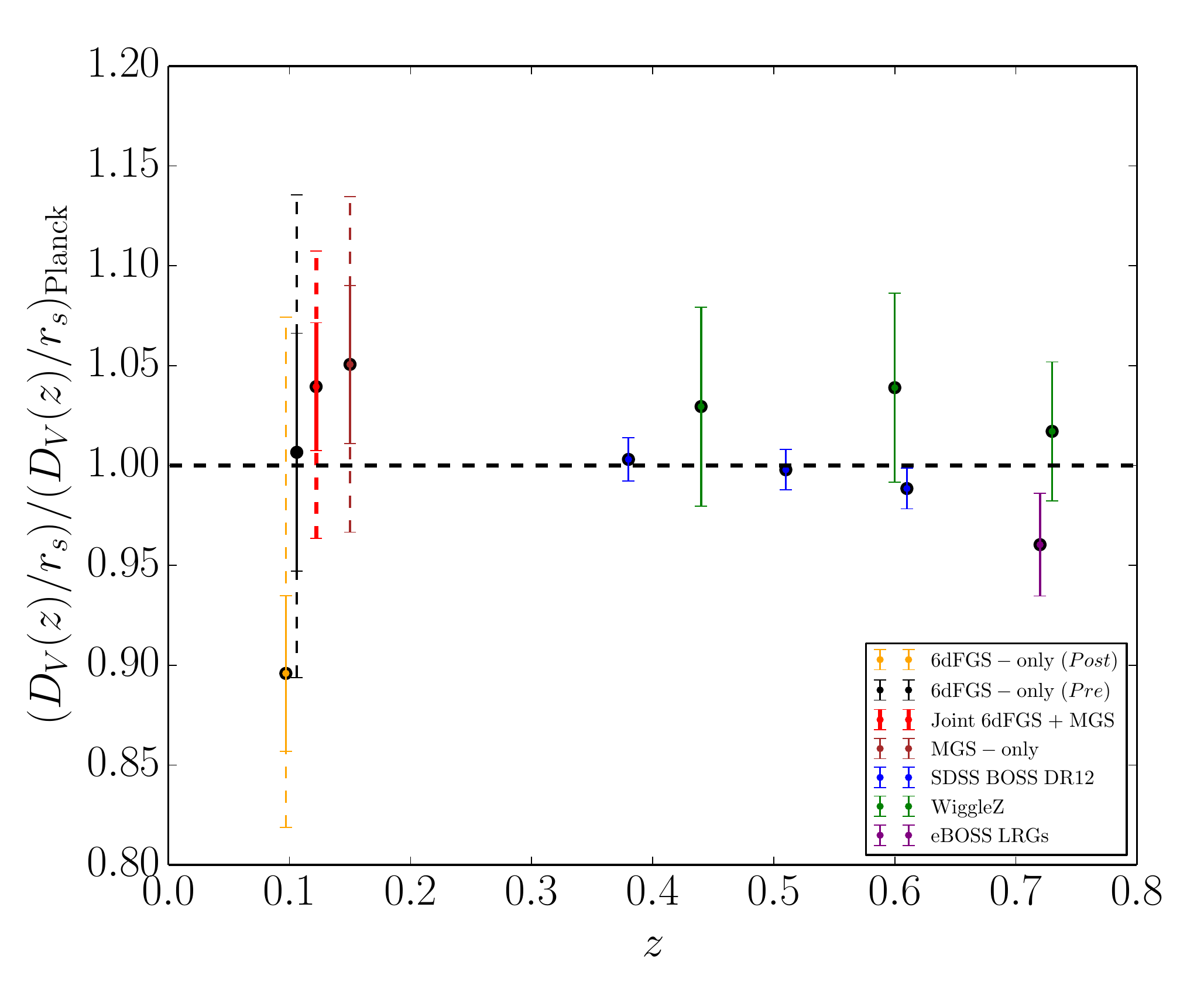}
    \caption{Distance constraints of  low redshift measurements BOSS DR12, eBOSS LRGs, WiggleZ and the joint 6dFGS/SDSS MGS constraint of $\alpha = 1.040 \pm 0.032$. For comparison purposes we focus on measurements at $z<0.8$, higher redshift measurements exist for both eBOSS QSO and BOSS Ly$\alpha$. The dashed line corresponds to a flat $\Lambda \mathrm{CDM}$ model matching the Planck best-fit. The pre-reconstruction 6dFGS result is from \citet{2011MNRAS.416.3017B} and the post-reconstruction is from this work. Error-bars are solid for $1\sigma$ and dashed for $2\sigma$ (focussing on the datasets used) it should be remembered that the post-reconstruction result from 6dFGS alone has a non-Gaussian likelihood such that the $\Lambda\mathrm{CDM}$ cosmology is consistent at the $2\sigma$ level.}
    \label{fig:figH}
\end{figure*}

\subsection{Combined Analysis with SDSS DR7 MGS}
\label{sec:joint}
The SDSS DR7 Main Galaxy Sample was analysed using the density field reconstruction technique in \cite{2015MNRAS.449..835R}, finding a 4\% distance measurement at $z_{\mathrm{eff}}=0.15$. By combining the likelihoods from SDSS MGS and 6dFGS post-reconstruction we consider the relative contributions of both surveys. As there is less than $3\%$ angular overlap between SDSS MGS and 6dFGS and the redshift distribution is sufficiently different, the covariance between surveys has a negligible contribution to the result.

A joint constraint between the post-reconstruction 6dFGS and SDSS MGS is obtained by multiplicatively combining the likelihoods, taking the consensus $\xi(s)+P(k)$ result for SDSS MGS available as a supplement from \cite{2015MNRAS.449..835R}, which alone gives $\alpha = 1.040 \pm 0.037$. When these likelihoods are combined a constraint of $\alpha = 1.040 \pm 0.032$ is obtained. The individual $\Delta\chi^{2}$ distributions are shown in Figure~\ref{fig:figJ}. Although the majority of the information comes from SDSS MGS, 6dFGS adds enough information to provide an improvement of $\sim 16\%$ on the SDSS MGS result. This combined result offers the most robust BAO constraint at low redshift to date.

Taking this joint constraint on $\alpha$ we can provide a distance measurement at the effective volume weighted redshift of joint 6dFGS and SDSS MGS, $z^{\mathrm{joint}}_{\mathrm{eff}}=0.122$, of $D_{V}(z_{\mathrm{eff}}=0.122)=539\pm17(r_{s}/r^{\mathrm{fid}}_{s})\,\mathrm{Mpc}$, with $D_{V}^{\mathrm{fid}}(z_{\mathrm{eff}}=0.122)=519\,\mathrm{Mpc}$. The joint post-reconstruction measurement of the distance is compared with the other survey results in Figure~\ref{fig:figH}, and also to the fiducial flat $\Lambda\mathrm{CDM}$ cosmology in the redshift range $z < 0.8$.

\subsection{Combining with the Planck 2015 $\Omega_{m}h^{2}$ prior}
\label{sec:8.3}
\begin{figure}
	\includegraphics[width=\columnwidth]{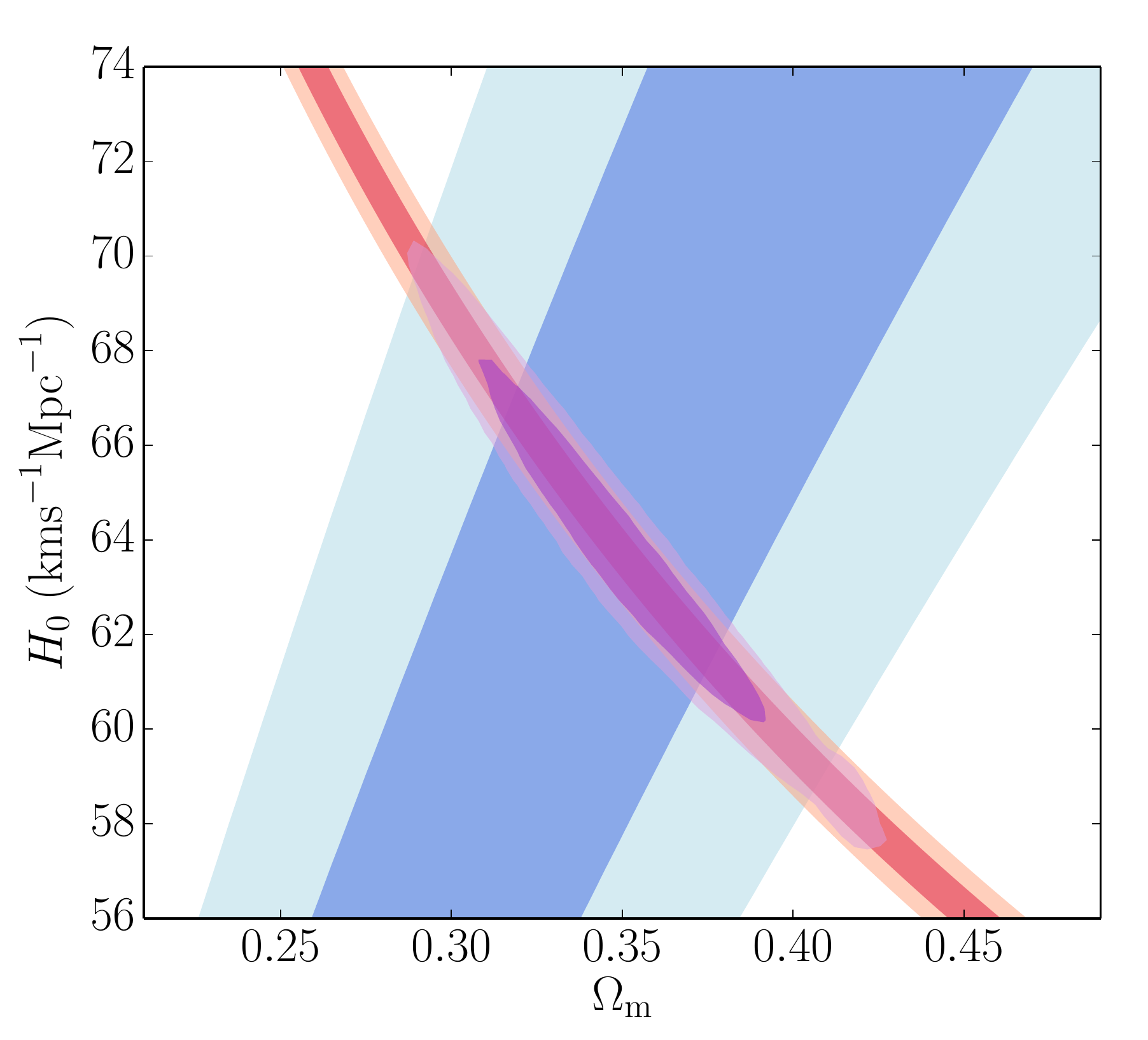}
    \caption{$H_{0}-\Omega_{m}$ parameter contours from the Planck 2015 $\Omega_{m}h^{2}=0.1417\pm0.024$ prior (red) and the $D_{v}/r_{s}=3.66\pm0.12$ (blue) measurements from the combined constraint of 6dFGS and SDSS MGS. Combining these individual constraints breaks the degeneracy to give a combined contour (purple) that allows for a measure of $H_{0}$ and $\Omega_{m}$. The corresponding results are given in Section \ref{sec:8.3} and Table \ref{tab:2}}
    \label{fig:figC}
\end{figure}

The ratio of $D_{V}/r_{s}$ can be constructed by using a numerically calibrated analytical approximation \citep{PhysRevD.92.123516} to CAMB of the sound horizon at the drag epoch (fiducial $r_{s}^{\mathrm{fid}}(z_{d})=147.5\,\mathrm{Mpc}$),

\begin{equation}
    r_{s} = 55.154\frac{\exp(-72.3(\Omega_\nu h^{2}+6\times10^{-4})^{2})}{(\Omega_{b}h^{2})^{0.12807}(\Omega_{m}h^{2}-\Omega_\nu h^{2})^{0.2535}},
    \label{eq:equ19}
\end{equation}

\noindent which can be propagated through to give a degenerate contour in $H_{0}-\Omega_{m}$ parameter space. In the above analytical expression $\Omega_{\nu}$ and $\Omega_{b}$ are the density parameters for neutrinos and baryons with values $1.4\times10^{-3}$ and $0.0484$ respectively. Taking the $\Omega_{m}h^{2}$ prior from the Planck 2015 results ($\Omega_{m}h^{2}=0.1417\pm0.024$) \citep{2015arXiv150201589P} the degeneracy can be broken to provide a measurement of $H_{0} = 64.0\pm3.5\,\mathrm{kms}^{-1}\mathrm{Mpc}$ and $\Omega_{m} = 0.346\pm0.045$. A plot showing the joint fit $H_{0}-\Omega_{m}$ contour with the Planck $\Omega_{m}h^{2}$ prior is given in Figure \ref{fig:figC}.

\section{Conclusion} 
\label{sec:concl}
We have updated the BAO analysis of 6dFGS presented in \cite{2011MNRAS.416.3017B}. Changes made to the analysis pipeline in comparison to that work include:
\begin{itemize}
\item Production of 600 COLA based fast mock catalogues over periodic boxes of size $1.2\,h^{-1}\mathrm{Gpc}$, populated with galaxies to describe the 6dFGS data sample through the application of a HOD modelling on top of a FoF halo finder. These robust mocks offer a more accurate covariance matrix with which to preform the analysis in comparison to the log-normal realisations generated in previous work.
\item A FFT version of density field reconstruction has been applied to the 6dFGS data and mock realisations.
\item We make use of a correlation function model that is now commonly used in SDSS (MGS and BOSS) BAO analyses. This model unlike the previous work, allows for freedom in the broadband shape by introducing polynomial terms to marginalise over. This is imperative when dealing with the post-reconstruction 2-point statistics as a well defined physical model for the broadband shape is still not known and it ensures that the information being used in the constraint is coming solely from the peak position.
\end{itemize}

Overall we find that, in the subsample of the mock catalogues that show a detection of the BAO peak both pre/post-reconstruction, $80\%$ show an improvement after the application of density field reconstruction to remove non-linear shifts of the galaxies, and the significance of detection improves on average. In the mean of the mock population an improvement factor of $I=\sigma_{\alpha, \mathrm{pre}}/\sigma_{\alpha, \mathrm{post}}\sim 1.86$ is seen. In the data catalogue, pre-reconstruction we find consistent results to the previous analysis of \cite{2011MNRAS.416.3017B}. In performing density field reconstruction to the data realisation we introduce a bi-modality into the likelihood that we argue is likely a statistical fluctuation against the mock population. At the $1\sigma$ confidence interval the error drops from $\sim 6.6\%$ to $\sim 4.1\%$, however due to the non-Gaussian nature of the post-reconstruction likelihood the significance of detection drops from $\sim 1.9\sigma$ to $\sim 1.8\sigma$. This decrease in statistical significance post-reconstruction is displayed in $28\%$ of the mock sample suggesting it is consistent with statistical fluctuations within the population. The measurements agree with the flat $\Lambda\mathrm{CDM}$ Planck based cosmology at the 2$\sigma$ level.

By combining the post-reconstruction result with that from SDSS MGS at $z_{\mathrm{eff}}=0.15$, we obtain a joint constraint of $\alpha(z_{\mathrm{eff}}=0.122) = 1.040 \pm 0.032$. The likelihood of 6dFGS offers a $\sim 16\%$ improvement on the SDSS MGS result at the 1$\sigma$ level. Using the $\Omega_{m}h^{2}$ prior from Planck 2015 results, the degeneracy of the joint 6dFGS and SDSS MGS $D_{V}/r_{s}$ constraint in the $\Omega_{m}-H_{0}$ plane is broken, as shown in Figure.~\ref{fig:figC}. This provides cosmological parameter constraints of $H_{0}=64.0\pm3.5\,\mathrm{kms}^{-1}\mathrm{Mpc}$ and $\Omega_{m}=0.346\pm0.045$. The final measurements of $\alpha$ are collated in Table \ref{tab:2}. While these results are consistent with the currently accepted Planck $\Lambda \mathrm{CDM}$ model, they are in $2.4\sigma$ tension with the latest Supernova Ia results ($H_{0}=73.24\pm1.74\,\mathrm{kms}^{-1}\mathrm{Mpc}$; \cite{2016ApJ...826...56R}).

This work constitutes the current best constraint available from BAO in the low redshift ($z < 0.3$) regime. In the near future both the Taipan Galaxy Survey \citep{2017PASA...34...47D} and DESI \citep{2016arXiv161100036D} will target this low redshift regime providing measurement of BAO at the $1\%$ level. These will further study any potential late time deviations from the currently accepted $\Lambda$CDM model.

\begin{table}
 \centering
 \caption{Summary table of results from the analysis.}
 \label{tab:2}
 \begin{tabular}{cccccc}
  \hline
  Result & $\alpha$ & $\chi^{2}/\nu$\\
  \hline
  6dF only - Pre-recon & $1.018\pm0.066(\pm^{0.179}_{0.116})$ & $0.81$\\
  6dF only - Post-recon & $0.895\pm0.041(\pm^{0.235}_{0.079})$ & $0.86$\\
  Joint Result - Post-recon & $1.040\pm0.032(\pm^{0.068}_{0.076})$ & $1.14$\\
  \hline
  Cosmological Parameters & (combined with Planck prior) & \\
  \hline
  $\Omega_{m}$ & $0.346\pm0.045$ & \\
  $H_{0}$ & $64.0\pm3.5\,\mathrm{kms}^{-1}\mathrm{Mpc}$ & \\
 \end{tabular}
\end{table}

\section*{Acknowledgements}

PC and WJP acknowledge support from the European Research Council through the Darksurvey grant 614030. WJP also acknowledges support from the UK Science and Technology Facilities Council grant ST/N000668/1 and the UK Space Agency grant ST/N00180X/1. FB is a Royal Society University Research Fellow. JK is supported by MUIR PRIN 2015 "Cosmology and Fundamental Physics: Illuminating the Dark Universe with Euclid" and Agenzia Spaziale Italiana agreement ASI/INAF/I/023/12/0. Parts of this research were conducted by the Australian Research Council Centre of Excellence for All-sky Astrophysics (CAASTRO), through project number CE110001020. The 6dF Galaxy Survey had many contributions to the instrument, survey and science, in particular we thank Matthew Colless, Heath Jones, Will Saunders, Fred Watson, Quentin Parker, Mike Read, Lachlan Campbell, Chris Springob, Christina Magoulas, John Lucey, Jeremy Mould, and Tom Jarrett, as well as the staff of the Australian Astronomical Observatory and other members of the 6dFGS team. This project has made use of the SCIAMA High Performance Computing (HPC) cluster at the ICG and also the GREEN-II Supercomputer in Swinburne. This work was supported by the Flagship Allocation Scheme of the NCI National Facility at the ANU.

\bibliographystyle{mnras}
\bibliography{referencesCarter18}

\begin{thebibliography}{}
\makeatletter
\relax
\def\mn@urlcharsother{\let\do\@makeother \do\$\do\&\do\#\do\^\do\_\do\%\do\~}
\def\mn@doi{\begingroup\mn@urlcharsother \@ifnextchar [ {\mn@doi@}
  {\mn@doi@[]}}
\def\mn@doi@[#1]#2{\def\@tempa{#1}\ifx\@tempa\@empty \href
  {http://dx.doi.org/#2} {doi:#2}\else \href {http://dx.doi.org/#2} {#1}\fi
  \endgroup}
\def\mn@eprint#1#2{\mn@eprint@#1:#2::\@nil}
\def\mn@eprint@arXiv#1{\href {http://arxiv.org/abs/#1} {{\tt arXiv:#1}}}
\def\mn@eprint@dblp#1{\href {http://dblp.uni-trier.de/rec/bibtex/#1.xml}
  {dblp:#1}}
\def\mn@eprint@#1:#2:#3:#4\@nil{\def\@tempa {#1}\def\@tempb {#2}\def\@tempc
  {#3}\ifx \@tempc \@empty \let \@tempc \@tempb \let \@tempb \@tempa \fi \ifx
  \@tempb \@empty \def\@tempb {arXiv}\fi \@ifundefined
  {mn@eprint@\@tempb}{\@tempb:\@tempc}{\expandafter \expandafter \csname
  mn@eprint@\@tempb\endcsname \expandafter{\@tempc}}}

\bibitem[\protect\citeauthoryear{{Achitouv}, {Blake}, {Carter}, {Koda}  \&
  {Beutler}}{{Achitouv} et~al.}{2017}]{2017PhRvD..95h3502A}
{Achitouv} I.,  {Blake} C.,  {Carter} P.,  {Koda} J.,   {Beutler} F.,  2017,
  \mn@doi [\prd] {10.1103/PhysRevD.95.083502}, \href
  {http://adsabs.harvard.edu/abs/2017PhRvD..95h3502A} {95, 083502}

\bibitem[\protect\citeauthoryear{{Alam} et~al.,}{{Alam}
  et~al.}{2017}]{2017MNRAS.470.2617A}
{Alam} S.,  et~al., 2017, \mn@doi [\mnras] {10.1093/mnras/stx721}, \href
  {http://adsabs.harvard.edu/abs/2017MNRAS.470.2617A} {470, 2617}

\bibitem[\protect\citeauthoryear{{Anderson} et~al.,}{{Anderson}
  et~al.}{2014}]{2014MNRAS.441...24A}
{Anderson} L.,  et~al., 2014, \mn@doi [\mnras] {10.1093/mnras/stu523}, \href
  {http://adsabs.harvard.edu/abs/2014MNRAS.441...24A} {441, 24}

\bibitem[\protect\citeauthoryear{{Ata} et~al.,}{{Ata}
  et~al.}{2018}]{2018MNRAS.473.4773A}
{Ata} M.,  et~al., 2018, \mn@doi [\mnras] {10.1093/mnras/stx2630}, \href
  {http://adsabs.harvard.edu/abs/2018MNRAS.473.4773A} {473, 4773}

\bibitem[\protect\citeauthoryear{Aubourg et~al.,}{Aubourg
  et~al.}{2015}]{PhysRevD.92.123516}
Aubourg E.,  et~al., 2015, \mn@doi [Phys. Rev. D] {10.1103/PhysRevD.92.123516},
  92, 123516

\bibitem[\protect\citeauthoryear{{Bautista} et~al.,}{{Bautista}
  et~al.}{2017}]{2017arXiv171208064B}
{Bautista} J.~E.,  et~al., 2017, preprint, \href
  {http://adsabs.harvard.edu/abs/2017arXiv171208064B} {} (\mn@eprint {arXiv}
  {1712.08064})

\bibitem[\protect\citeauthoryear{{Beutler} et~al.,}{{Beutler}
  et~al.}{2011}]{2011MNRAS.416.3017B}
{Beutler} F.,  et~al., 2011, \mn@doi [\mnras]
  {10.1111/j.1365-2966.2011.19250.x}, \href
  {http://adsabs.harvard.edu/abs/2011MNRAS.416.3017B} {416, 3017}

\bibitem[\protect\citeauthoryear{{Beutler} et~al.,}{{Beutler}
  et~al.}{2013}]{2013MNRAS.429.3604B}
{Beutler} F.,  et~al., 2013, \mn@doi [\mnras] {10.1093/mnras/sts637}, \href
  {http://adsabs.harvard.edu/abs/2013MNRAS.429.3604B} {429, 3604}

\bibitem[\protect\citeauthoryear{{Bianchi} \& {Percival}}{{Bianchi} \&
  {Percival}}{2017}]{2017MNRAS.472.1106B}
{Bianchi} D.,  {Percival} W.~J.,  2017, \mn@doi [\mnras]
  {10.1093/mnras/stx2053}, \href
  {http://adsabs.harvard.edu/abs/2017MNRAS.472.1106B} {472, 1106}

\bibitem[\protect\citeauthoryear{{Blake} et~al.,}{{Blake}
  et~al.}{2011}]{2011MNRAS.415.2892B}
{Blake} C.,  et~al., 2011, \mn@doi [\mnras] {10.1111/j.1365-2966.2011.19077.x},
  \href {http://adsabs.harvard.edu/abs/2011MNRAS.415.2892B} {415, 2892}

\bibitem[\protect\citeauthoryear{{Blake}, {Carter}  \& {Koda}}{{Blake}
  et~al.}{2018}]{2018arXiv180104969B}
{Blake} C.,  {Carter} P.,   {Koda} J.,  2018, preprint, \href
  {http://adsabs.harvard.edu/abs/2018arXiv180104969B} {} (\mn@eprint {arXiv}
  {1801.04969})

\bibitem[\protect\citeauthoryear{{Bullock}, {Kolatt}, {Sigad}, {Somerville},
  {Kravtsov}, {Klypin}, {Primack}  \& {Dekel}}{{Bullock}
  et~al.}{2001}]{2001MNRAS.321..559B}
{Bullock} J.~S.,  {Kolatt} T.~S.,  {Sigad} Y.,  {Somerville} R.~S.,  {Kravtsov}
  A.~V.,  {Klypin} A.~A.,  {Primack} J.~R.,   {Dekel} A.,  2001, \mn@doi
  [\mnras] {10.1046/j.1365-8711.2001.04068.x}, \href
  {http://adsabs.harvard.edu/abs/2001MNRAS.321..559B} {321, 559}

\bibitem[\protect\citeauthoryear{{Burden}, {Percival}, {Manera}, {Cuesta},
  {Vargas Magana}  \& {Ho}}{{Burden} et~al.}{2014}]{2014MNRAS.445.3152B}
{Burden} A.,  {Percival} W.~J.,  {Manera} M.,  {Cuesta} A.~J.,  {Vargas Magana}
  M.,   {Ho} S.,  2014, \mn@doi [\mnras] {10.1093/mnras/stu1965}, \href
  {http://adsabs.harvard.edu/abs/2014MNRAS.445.3152B} {445, 3152}

\bibitem[\protect\citeauthoryear{{Burden}, {Percival}  \& {Howlett}}{{Burden}
  et~al.}{2015}]{2015MNRAS.453..456B}
{Burden} A.,  {Percival} W.~J.,   {Howlett} C.,  2015, \mn@doi [\mnras]
  {10.1093/mnras/stv1581}, \href
  {http://adsabs.harvard.edu/abs/2015MNRAS.453..456B} {453, 456}

\bibitem[\protect\citeauthoryear{{Cole} et~al.,}{{Cole}
  et~al.}{2005}]{2005MNRAS.362..505C}
{Cole} S.,  et~al., 2005, \mn@doi [\mnras] {10.1111/j.1365-2966.2005.09318.x},
  \href {http://adsabs.harvard.edu/abs/2005MNRAS.362..505C} {362, 505}

\bibitem[\protect\citeauthoryear{{Colless} et~al.,}{{Colless}
  et~al.}{2001}]{2001MNRAS.328.1039C}
{Colless} M.,  et~al., 2001, \mn@doi [\mnras]
  {10.1046/j.1365-8711.2001.04902.x}, \href
  {http://adsabs.harvard.edu/abs/2001MNRAS.328.1039C} {328, 1039}

\bibitem[\protect\citeauthoryear{{Crocce} \& {Scoccimarro}}{{Crocce} \&
  {Scoccimarro}}{2008}]{2008PhRvD..77b3533C}
{Crocce} M.,  {Scoccimarro} R.,  2008, \mn@doi [\prd]
  {10.1103/PhysRevD.77.023533}, \href
  {http://adsabs.harvard.edu/abs/2008PhRvD..77b3533C} {77, 023533}

\bibitem[\protect\citeauthoryear{{DESI Collaboration} et~al.,}{{DESI
  Collaboration} et~al.}{2016}]{2016arXiv161100036D}
{DESI Collaboration} et~al., 2016, preprint, \href
  {http://adsabs.harvard.edu/abs/2016arXiv161100036D} {} (\mn@eprint {arXiv}
  {1611.00036})

\bibitem[\protect\citeauthoryear{{Delubac} et~al.,}{{Delubac}
  et~al.}{2015}]{2015A&A...574A..59D}
{Delubac} T.,  et~al., 2015, \mn@doi [\aap] {10.1051/0004-6361/201423969},
  \href {http://adsabs.harvard.edu/abs/2015A%26A...574A..59D} {574, A59}

\bibitem[\protect\citeauthoryear{{Eisenstein} \& {Hu}}{{Eisenstein} \&
  {Hu}}{1998}]{1998ApJ...496..605E}
{Eisenstein} D.~J.,  {Hu} W.,  1998, \mn@doi [\apj] {10.1086/305424}, \href
  {http://adsabs.harvard.edu/abs/1998ApJ...496..605E} {496, 605}

\bibitem[\protect\citeauthoryear{{Eisenstein} et~al.,}{{Eisenstein}
  et~al.}{2005}]{2005ApJ...633..560E}
{Eisenstein} D.~J.,  et~al., 2005, \mn@doi [\apj] {10.1086/466512}, \href
  {http://adsabs.harvard.edu/abs/2005ApJ...633..560E} {633, 560}

\bibitem[\protect\citeauthoryear{{Eisenstein}, {Seo}  \& {White}}{{Eisenstein}
  et~al.}{2007a}]{2007ApJ...664..660E}
{Eisenstein} D.~J.,  {Seo} H.-J.,   {White} M.,  2007a, \mn@doi [\apj]
  {10.1086/518755}, \href {http://adsabs.harvard.edu/abs/2007ApJ...664..660E}
  {664, 660}

\bibitem[\protect\citeauthoryear{{Eisenstein}, {Seo}, {Sirko}  \&
  {Spergel}}{{Eisenstein} et~al.}{2007b}]{2007ApJ...664..675E}
{Eisenstein} D.~J.,  {Seo} H.-J.,  {Sirko} E.,   {Spergel} D.~N.,  2007b,
  \mn@doi [\apj] {10.1086/518712}, \href
  {http://adsabs.harvard.edu/abs/2007ApJ...664..675E} {664, 675}

\bibitem[\protect\citeauthoryear{{Feldman}, {Kaiser}  \& {Peacock}}{{Feldman}
  et~al.}{1994}]{1994ApJ...426...23F}
{Feldman} H.~A.,  {Kaiser} N.,   {Peacock} J.~A.,  1994, \mn@doi [\apj]
  {10.1086/174036}, \href {http://adsabs.harvard.edu/abs/1994ApJ...426...23F}
  {426, 23}

\bibitem[\protect\citeauthoryear{{Font-Ribera} et~al.,}{{Font-Ribera}
  et~al.}{2014}]{2014JCAP...05..027F}
{Font-Ribera} A.,  et~al., 2014, \mn@doi [\jcap]
  {10.1088/1475-7516/2014/05/027}, \href
  {http://adsabs.harvard.edu/abs/2014JCAP...05..027F} {5, 027}

\bibitem[\protect\citeauthoryear{{Hartlap}, {Simon}  \& {Schneider}}{{Hartlap}
  et~al.}{2007}]{2007A&A...464..399H}
{Hartlap} J.,  {Simon} P.,   {Schneider} P.,  2007, \mn@doi [\aap]
  {10.1051/0004-6361:20066170}, \href
  {http://adsabs.harvard.edu/abs/2007A%26A...464..399H} {464, 399}

\bibitem[\protect\citeauthoryear{{Howlett}, {Manera}  \& {Percival}}{{Howlett}
  et~al.}{2015}]{2015A&C....12..109H}
{Howlett} C.,  {Manera} M.,   {Percival} W.~J.,  2015, \mn@doi [Astronomy and
  Computing] {10.1016/j.ascom.2015.07.003}, \href
  {http://adsabs.harvard.edu/abs/2015A%26C....12..109H} {12, 109}

\bibitem[\protect\citeauthoryear{{Jones} et~al.,}{{Jones}
  et~al.}{2004}]{2004MNRAS.355..747J}
{Jones} D.~H.,  et~al., 2004, \mn@doi [\mnras]
  {10.1111/j.1365-2966.2004.08353.x}, \href
  {http://adsabs.harvard.edu/abs/2004MNRAS.355..747J} {355, 747}

\bibitem[\protect\citeauthoryear{{Jones}, {Peterson}, {Colless}  \&
  {Saunders}}{{Jones} et~al.}{2006}]{2006MNRAS.369...25J}
{Jones} D.~H.,  {Peterson} B.~A.,  {Colless} M.,   {Saunders} W.,  2006,
  \mn@doi [\mnras] {10.1111/j.1365-2966.2006.10291.x}, \href
  {http://adsabs.harvard.edu/abs/2006MNRAS.369...25J} {369, 25}

\bibitem[\protect\citeauthoryear{{Jones} et~al.,}{{Jones}
  et~al.}{2009}]{2009MNRAS.399..683J}
{Jones} D.~H.,  et~al., 2009, \mn@doi [\mnras]
  {10.1111/j.1365-2966.2009.15338.x}, \href
  {http://adsabs.harvard.edu/abs/2009MNRAS.399..683J} {399, 683}

\bibitem[\protect\citeauthoryear{{Kaiser}}{{Kaiser}}{1987}]{1987MNRAS.227....1K}
{Kaiser} N.,  1987, \mn@doi [\mnras] {10.1093/mnras/227.1.1}, \href
  {http://adsabs.harvard.edu/abs/1987MNRAS.227....1K} {227, 1}

\bibitem[\protect\citeauthoryear{{Kazin} et~al.,}{{Kazin}
  et~al.}{2010}]{2010ApJ...710.1444K}
{Kazin} E.~A.,  et~al., 2010, \mn@doi [\apj] {10.1088/0004-637X/710/2/1444},
  \href {http://adsabs.harvard.edu/abs/2010ApJ...710.1444K} {710, 1444}

\bibitem[\protect\citeauthoryear{{Kazin} et~al.,}{{Kazin}
  et~al.}{2014}]{2014MNRAS.441.3524K}
{Kazin} E.~A.,  et~al., 2014, \mn@doi [\mnras] {10.1093/mnras/stu778}, \href
  {http://adsabs.harvard.edu/abs/2014MNRAS.441.3524K} {441, 3524}

\bibitem[\protect\citeauthoryear{{Koda}, {Blake}, {Beutler}, {Kazin}  \&
  {Marin}}{{Koda} et~al.}{2016}]{2016MNRAS.459.2118K}
{Koda} J.,  {Blake} C.,  {Beutler} F.,  {Kazin} E.,   {Marin} F.,  2016,
  \mn@doi [\mnras] {10.1093/mnras/stw763}, \href
  {http://adsabs.harvard.edu/abs/2016MNRAS.459.2118K} {459, 2118}

\bibitem[\protect\citeauthoryear{{Landy} \& {Szalay}}{{Landy} \&
  {Szalay}}{1993}]{1993ApJ...412...64L}
{Landy} S.~D.,  {Szalay} A.~S.,  1993, \mn@doi [\apj] {10.1086/172900}, \href
  {http://adsabs.harvard.edu/abs/1993ApJ...412...64L} {412, 64}

\bibitem[\protect\citeauthoryear{Lewis, Challinor  \& Lasenby}{Lewis
  et~al.}{2000}]{Lewis:1999bs}
Lewis A.,  Challinor A.,   Lasenby A.,  2000, Astrophys. J., 538, 473

\bibitem[\protect\citeauthoryear{{Navarro}, {Frenk}  \& {White}}{{Navarro}
  et~al.}{1996}]{1996ApJ...462..563N}
{Navarro} J.~F.,  {Frenk} C.~S.,   {White} S.~D.~M.,  1996, \mn@doi [\apj]
  {10.1086/177173}, \href {http://adsabs.harvard.edu/abs/1996ApJ...462..563N}
  {462, 563}

\bibitem[\protect\citeauthoryear{{Noh}, {White}  \& {Padmanabhan}}{{Noh}
  et~al.}{2009}]{2009PhRvD..80l3501N}
{Noh} Y.,  {White} M.,   {Padmanabhan} N.,  2009, \mn@doi [\prd]
  {10.1103/PhysRevD.80.123501}, \href
  {http://adsabs.harvard.edu/abs/2009PhRvD..80l3501N} {80, 123501}

\bibitem[\protect\citeauthoryear{{Padmanabhan}, {White}  \&
  {Cohn}}{{Padmanabhan} et~al.}{2009}]{2009PhRvD..79f3523P}
{Padmanabhan} N.,  {White} M.,   {Cohn} J.~D.,  2009, \mn@doi [\prd]
  {10.1103/PhysRevD.79.063523}, \href
  {http://adsabs.harvard.edu/abs/2009PhRvD..79f3523P} {79, 063523}

\bibitem[\protect\citeauthoryear{{Padmanabhan}, {Xu}, {Eisenstein}, {Scalzo},
  {Cuesta}, {Mehta}  \& {Kazin}}{{Padmanabhan}
  et~al.}{2012}]{2012MNRAS.427.2132P}
{Padmanabhan} N.,  {Xu} X.,  {Eisenstein} D.~J.,  {Scalzo} R.,  {Cuesta} A.~J.,
   {Mehta} K.~T.,   {Kazin} E.,  2012, \mn@doi [\mnras]
  {10.1111/j.1365-2966.2012.21888.x}, \href
  {http://adsabs.harvard.edu/abs/2012MNRAS.427.2132P} {427, 2132}

\bibitem[\protect\citeauthoryear{{Percival} et~al.,}{{Percival}
  et~al.}{2001}]{2001MNRAS.327.1297P}
{Percival} W.~J.,  et~al., 2001, \mn@doi [\mnras]
  {10.1046/j.1365-8711.2001.04827.x}, \href
  {http://adsabs.harvard.edu/abs/2001MNRAS.327.1297P} {327, 1297}

\bibitem[\protect\citeauthoryear{{Percival} et~al.,}{{Percival}
  et~al.}{2010}]{2010MNRAS.401.2148P}
{Percival} W.~J.,  et~al., 2010, \mn@doi [\mnras]
  {10.1111/j.1365-2966.2009.15812.x}, \href
  {http://adsabs.harvard.edu/abs/2010MNRAS.401.2148P} {401, 2148}

\bibitem[\protect\citeauthoryear{{Planck Collaboration} et~al.,}{{Planck
  Collaboration} et~al.}{2015}]{2015arXiv150201589P}
{Planck Collaboration} et~al., 2015, preprint, \href
  {http://adsabs.harvard.edu/abs/2015arXiv150201589P} {} (\mn@eprint {arXiv}
  {1502.01589})

\bibitem[\protect\citeauthoryear{{Riess} et~al.,}{{Riess}
  et~al.}{2016}]{2016ApJ...826...56R}
{Riess} A.~G.,  et~al., 2016, \mn@doi [\apj] {10.3847/0004-637X/826/1/56},
  \href {http://adsabs.harvard.edu/abs/2016ApJ...826...56R} {826, 56}

\bibitem[\protect\citeauthoryear{{Ross}, {Samushia}, {Howlett}, {Percival},
  {Burden}  \& {Manera}}{{Ross} et~al.}{2015}]{2015MNRAS.449..835R}
{Ross} A.~J.,  {Samushia} L.,  {Howlett} C.,  {Percival} W.~J.,  {Burden} A.,
  {Manera} M.,  2015, \mn@doi [\mnras] {10.1093/mnras/stv154}, \href
  {http://adsabs.harvard.edu/abs/2015MNRAS.449..835R} {449, 835}

\bibitem[\protect\citeauthoryear{{Schmittfull}, {Feng}, {Beutler}, {Sherwin}
  \& {Chu}}{{Schmittfull} et~al.}{2015}]{2015PhRvD..92l3522S}
{Schmittfull} M.,  {Feng} Y.,  {Beutler} F.,  {Sherwin} B.,   {Chu} M.~Y.,
  2015, \mn@doi [\prd] {10.1103/PhysRevD.92.123522}, \href
  {http://adsabs.harvard.edu/abs/2015PhRvD..92l3522S} {92, 123522}

\bibitem[\protect\citeauthoryear{{Seo} \& {Eisenstein}}{{Seo} \&
  {Eisenstein}}{2007}]{2007ApJ...665...14S}
{Seo} H.-J.,  {Eisenstein} D.~J.,  2007, \mn@doi [\apj] {10.1086/519549}, \href
  {http://adsabs.harvard.edu/abs/2007ApJ...665...14S} {665, 14}

\bibitem[\protect\citeauthoryear{{Slosar} et~al.,}{{Slosar}
  et~al.}{2013}]{2013JCAP...04..026S}
{Slosar} A.,  et~al., 2013, \mn@doi [\jcap] {10.1088/1475-7516/2013/04/026},
  \href {http://adsabs.harvard.edu/abs/2013JCAP...04..026S} {4, 026}

\bibitem[\protect\citeauthoryear{{Tassev}, {Zaldarriaga}  \&
  {Eisenstein}}{{Tassev} et~al.}{2013}]{2013JCAP...06..036T}
{Tassev} S.,  {Zaldarriaga} M.,   {Eisenstein} D.~J.,  2013, \mn@doi [\jcap]
  {10.1088/1475-7516/2013/06/036}, \href
  {http://adsabs.harvard.edu/abs/2013JCAP...06..036T} {6, 036}

\bibitem[\protect\citeauthoryear{{York} et~al.,}{{York}
  et~al.}{2000}]{2000AJ....120.1579Y}
{York} D.~G.,  et~al., 2000, \mn@doi [\aj] {10.1086/301513}, \href
  {http://adsabs.harvard.edu/abs/2000AJ....120.1579Y} {120, 1579}

\bibitem[\protect\citeauthoryear{{Zehavi} et~al.,}{{Zehavi}
  et~al.}{2011}]{2011ApJ...736...59Z}
{Zehavi} I.,  et~al., 2011, \mn@doi [\apj] {10.1088/0004-637X/736/1/59}, \href
  {http://adsabs.harvard.edu/abs/2011ApJ...736...59Z} {736, 59}

\bibitem[\protect\citeauthoryear{{Zel'dovich}}{{Zel'dovich}}{1970}]{1970A&A.....5...84Z}
{Zel'dovich} Y.~B.,  1970, \aap, \href
  {http://adsabs.harvard.edu/abs/1970A%26A.....5...84Z} {5, 84}

\bibitem[\protect\citeauthoryear{{Zheng} et~al.,}{{Zheng}
  et~al.}{2005}]{2005ApJ...633..791Z}
{Zheng} Z.,  et~al., 2005, \mn@doi [\apj] {10.1086/466510}, \href
  {http://adsabs.harvard.edu/abs/2005ApJ...633..791Z} {633, 791}

\bibitem[\protect\citeauthoryear{{da Cunha} et~al.,}{{da Cunha}
  et~al.}{2017}]{2017PASA...34...47D}
{da Cunha} E.,  et~al., 2017, \mn@doi [\pasa] {10.1017/pasa.2017.41}, \href
  {http://adsabs.harvard.edu/abs/2017PASA...34...47D} {34, e047}

\bibitem[\protect\citeauthoryear{{de la Torre} et~al.,}{{de la Torre}
  et~al.}{2013}]{2013A&A...557A..54D}
{de la Torre} S.,  et~al., 2013, \mn@doi [\aap] {10.1051/0004-6361/201321463},
  \href {http://adsabs.harvard.edu/abs/2013A%26A...557A..54D} {557, A54}

\bibitem[\protect\citeauthoryear{{van den Bosch}, {Norberg}, {Mo}  \&
  {Yang}}{{van den Bosch} et~al.}{2004}]{2004MNRAS.352.1302V}
{van den Bosch} F.~C.,  {Norberg} P.,  {Mo} H.~J.,   {Yang} X.,  2004, \mn@doi
  [\mnras] {10.1111/j.1365-2966.2004.08021.x}, \href
  {http://adsabs.harvard.edu/abs/2004MNRAS.352.1302V} {352, 1302}

\makeatother
\end{thebibliography}

\bsp
\label{lastpage}
\end{document}